\documentclass[aps, pra,  10pt, twocolumn, superscriptaddress, longbibliography, floatfix, titlepage=false]{revtex4-1}
\usepackage[nointlimits]{amsmath}
\usepackage{amsthm}
\usepackage{amssymb, cancel}
\usepackage{graphicx, nicefrac}
\usepackage{color}
\usepackage[dvipsnames]{xcolor}
\usepackage{txfonts}
\usepackage{mathtools}
\usepackage{hyperref}
\usepackage{multirow}
\usepackage{tikz}
\hypersetup{colorlinks=true, citecolor=blue, linkcolor=blue}
\graphicspath{{./figures/}}

\usepackage[skip=4pt, indent=15pt]{parskip}
\usepackage{url}
\usepackage{todonotes}
\renewcommand{\selectlanguage}[1]{}

\DeclareMathOperator\erf{erf}
\newcommand\E[1]{{\mathbb{E}\left[#1\right]}}
\newcommand\V[1]{{\mathbb{V}\left[#1\right]}}

\newcommand\Cov[1]{{\text{Cov}\left[#1\right]}}
 
\begin{document}
\title{Featuremetric benchmarking: Quantum computer benchmarks based on circuit features}
\author{Timothy Proctor}
\affiliation{Quantum Performance Laboratory, Sandia National Laboratories, Livermore, CA 94550, USA}
\author{Anh Tran}
\affiliation{Center for Computing Research, Sandia National Laboratories, Albuquerque, NM 87123}
\author{Xingxin Liu}
\affiliation{Joint Quantum Institute, University of Maryland, College Park, MD 20742, USA}
\author{Aditya Dhumuntarao}
\author{Stefan Seritan}
\affiliation{Quantum Performance Laboratory, Sandia National Laboratories, Livermore, CA 94550, USA}
\author{Alaina Green}
\affiliation{Joint Quantum Institute, University of Maryland, College Park, MD 20742, USA}
\affiliation{National Quantum Laboratory (QLab), University of Maryland, College Park, MD 20742 USA}
\author{Norbert M Linke}
\affiliation{Joint Quantum Institute, University of Maryland, College Park, MD 20742, USA}
\affiliation{National Quantum Laboratory (QLab), University of Maryland, College Park, MD 20742 USA}
\affiliation{Duke Quantum Center, Duke University, Durham, NC 27701, USA}
\date{\today}
\begin{abstract} 
Benchmarks that concisely summarize the performance of many-qubit quantum computers are essential for measuring progress towards the goal of useful quantum computation. In this work, we present a benchmarking framework that is based on quantifying how a quantum computer's performance on quantum circuits varies as a function of \emph{features} of those circuits, such as circuit depth, width, two-qubit gate density, problem input size, or algorithmic depth. Our \emph{featuremetric benchmarking} framework generalizes volumetric benchmarking---a widely-used methodology that quantifies performance versus circuit width and depth---and we show that it enables richer and more faithful models of quantum computer performance. We demonstrate featuremetric benchmarking with example benchmarks run on IBM Q and IonQ systems of up to 27 qubits, and we show how to produce performance summaries from the data using Gaussian process regression. Our data analysis methods are also of interest in the special case of volumetric benchmarking, as they enable the creation of intuitive two-dimensional \emph{capability regions} using data from few circuits.
\end{abstract}
\maketitle

\section{Introduction}
Over the last decade, quantum computing hardware has advanced from few-qubit physics experiments \cite{Barends2014-ya} to 20-400 qubit prototypes of utility-scale quantum computers \cite{Arute2019-mk, Moses2023-do, Chen2023-la, Google-Quantum-AI-and-Collaborators2024-hq, Bluvstein2023-dp, Kim2023-si}. Contemporary quantum computers can outperform classical computers on specially designed tasks  \cite{Arute2019-mk} and, recently, many of the building-blocks needed for useful quantum computations have been demonstrated \cite{Google-Quantum-AI-and-Collaborators2024-hq, Bluvstein2023-dp, Krinner2022-tp, Google_Quantum_AI2023-yr, Gupta2024-pr}. These rapid advances in quantum computing hardware have been reflected in the development of a wide range of benchmarks for measuring system performance \cite{Proctor2024-av,  Hashim2024-om}. Quantum computers are complex integrated devices, and this complexity necessitates many distinct and complementary kinds of benchmark, each measuring different aspects of quantum computer performance \cite{Proctor2024-av}. For example, benchmarks can measure the error rates of one- and two-qubit gates \cite{Emerson2005-fd, Emerson2007-am, Knill2008-jf, Magesan2011-hc,  Magesan2012-dz}, the performance of its circuit compilers \cite{Kharkov2022-yw, Singh2023-in}, or the time taken to complete an algorithm \cite{Lubinski2023-mr}. 

Quantum computers obtain their computational power by executing quantum circuits with sufficiently low error, and so benchmarks that measure circuit execution capabilities are a particularly useful way to summarize quantum computer performance \cite{Proctor2024-av, Blume-Kohout2020-de, Proctor2021-wt}. Prominent examples include many application- and algorithm-based benchmarks \cite{Chen2022-dm, Linke2017-mr, Wright2019-zj, Tomesh2022-nu, Murali2019-my, Donkers2022-wt, Finzgar2022-aa, Mills2020-zh, Lubinski2023-zy, Lubinski2024-ci, Lubinski2023-mr, Chen2023-la, Benedetti2019-pp, Li2020-ry, Quetschlich2023-bg, Dong2021-gj, Martiel2021-vp, Van_der_Schoot2022-gv,Van_der_Schoot2023-vo, Cornelissen2021-yt, Georgopoulos2021-hh, Dong2022-ga}, mirror circuit benchmarks \cite{Hines2024-ae, Proctor2021-wt}, and the quantum volume benchmark \cite{Cross2019-ku, Corcoles2020-vn, Jurcevic2021-dz}. These circuit-level benchmarks directly capture the impact of many complex kinds of error that benchmarks for individual components (e.g., one- and two-qubit gates) are insensitive to \cite{Proctor2019-gf, Hines2023-tz, McKay2023-bx, Proctor2022-yl, Hines2023-vq, Hothem2023-pc}, including crosstalk \cite{Sarovar2020-pz, Gambetta2012-zd} and some non-Markovian errors \cite{Hothem2023-pc}. However, whereas most component-level benchmarks measure well-understood quantities (e.g., gate or layer fidelities \cite{Hashim2024-om, Emerson2005-fd, Emerson2007-am, Knill2008-jf, Magesan2011-hc,  Magesan2012-dz, Proctor2019-gf, Hines2023-tz, McKay2023-bx, Proctor2022-yl, Hines2023-vq, Helsen2019-cp, Helsen2022-yp}), from which performance predictions can be made \cite{Proctor2021-wt}, circuit-level benchmarks simply measure performance on some, necessarily small, set of circuits. It is often unclear how to predict the performance of other circuits from those results---or, in some cases, even how to provide insightful performance summaries from the data. 

\begin{figure}[t!]
    \centering  
    \includegraphics[width=7.5cm]{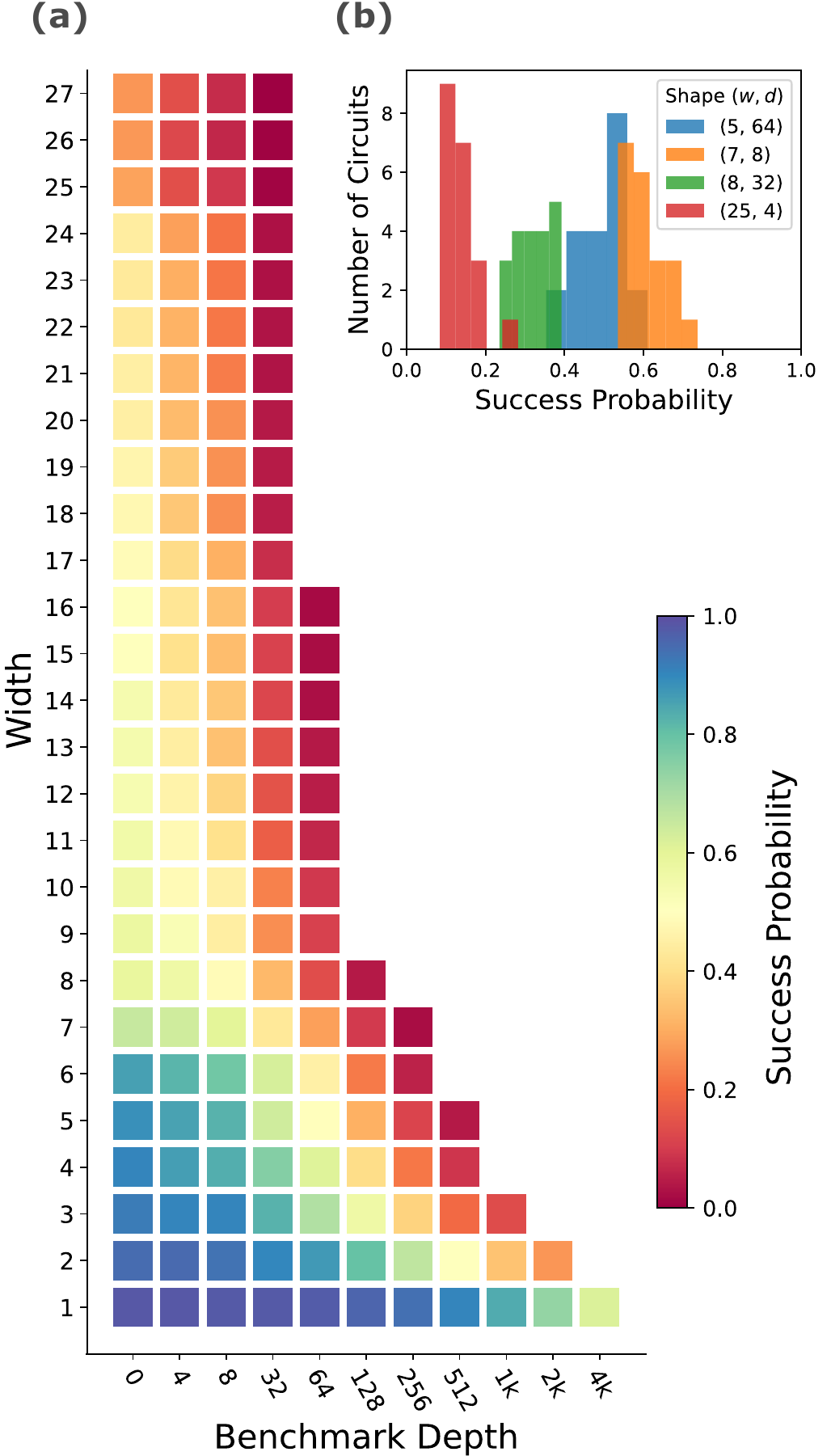}
    \caption{\textbf{Volumetric benchmarking of \texttt{ibmq\_montreal}}. \textbf{(a)} The results of a volumetric benchmark run on a 27-qubit IBM quantum computer (\texttt{ibmq\_montreal}). This plot shows  the mean success probability of randomized mirror circuits versus circuit shape (circuit width $w$ and benchmark depth  \cite{Proctor2022-yl} $d$). \textbf{(b)} Histograms of the success probabilities of the 20 circuits of each shape that were run, for a selection of circuit shapes. Each circuit's success probability was estimated from 1024 executions of that circuit, so the differences seen here are statistically significant. This demonstrates that, although the volumetric benchmarking plot of (a) shows mean performance as a function of circuit width and depth, two circuits with the same width and depth can have significantly different success probabilities.}
    \label{fig:ibm_montreal_vb}
\end{figure}

Volumetric benchmarking \cite{Blume-Kohout2020-de, Proctor2021-wt} is a popular approach \cite{Proctor2021-wt, Lubinski2023-zy, Lubinski2024-ci, Lubinski2023-mr, Chen2023-la} to creating systematic circuit-level benchmarks that are amenable to simple performance summaries. In volumetric benchmarking, a quantum computer's capability on some circuit family is mapped out as a function of circuit width ($w$, the number of qubits) and circuit depth ($d$). By measuring performance versus circuit shape $(w,d)$, the data can be simply plotted in the width $\times$ depth plane---as shown in the example of Fig.~\ref{fig:ibm_montreal_vb}---producing easily-understood summaries of performance. However, although volumetric benchmarking is a simple framework for creating quantum computer benchmarks and presenting their results, it has some important limitations.

Volumetric benchmarking designates two \emph{features} of circuits---their width and depth---as the variables against which circuit execution capabilities are measured. However, these two features are not always the most salient for understanding performance. This is demonstrated by the adaptation of volumetric benchmarking to, e.g., plot performance versus the number of two qubit gates (instead of depth) \cite{Chen2023-la} and the input problem size for a quantum algorithm (instead of width) \cite{Lubinski2023-mr}. Other features might be used instead of width or depth because they are expected to be more predictive of circuit performance, or simply because it is of intrinsic interest to understand how performance varies with these features. Furthermore, measuring performance versus only two features is motivated primarily by the convenience with which that data can be visualized (e.g., see Fig.~\ref{fig:ibm_montreal_vb}), and it is well-known that the performance of  contemporary quantum computers does not depend only on circuit width and depth---or any two features of circuits. 

No small set of circuit features is likely to be entirely predictive of overall execution error, but it is plausible that a small set of features will explain the majority of the variation in circuit error, enabling concise yet predictive summaries of performance. This could be explored with a general feature-based approach to quantum computer benchmarking, in which performance versus $\chi$-many different features is mapped out, and we introduce such a framework herein.

Extending volumetric benchmarking to a general feature-based approach to designing quantum computer benchmarks requires new ways to summarize those benchmark results. In volumetric benchmarking, data are simply plotted in the width $\times$ depth plane (and sometimes simplified into quantities such as binary ``capability regions'' \cite{Proctor2024-av, Proctor2021-wt, Hashim2024-om}). This has limitations even in the setting of two features, because it requires running many circuits to create performance summaries like that of Fig.~\ref{fig:ibm_montreal_vb}. But it fails for $\chi \geq 3$ features, as it is not feasible to measure performance at each feature value in even a moderately dense $\chi$-dimensional grid. Furthermore, it is not possible to summarize the results by simply plotting the data in a $\chi$ dimensional space.

In this paper, we introduce a general framework for circuit-feature-based benchmarking of quantum computers and we propose methods for creating performance summaries from these benchmarks' data. Our first contribution is a foundational framework---\emph{featuremetric benchmarking}---for how to benchmark quantum computer performance as a function of circuit features, together with applying this framework to benchmark IBM Q and IonQ systems using simple examples. We suggest some possible circuit features to use, but we do not attempt to provide a comprehensive list.

Our second contribution is to define the featuremetric benchmarking data analysis problem, and to explore one method for addressing it. This also provides new insights for the special case of volumetric benchmarking, i.e. the case of two features. The featuremetric benchmarking data analysis problem is an interpolation and extrapolation problem in $\chi$ dimensions, which is a ubiquitous problem in science. There are many machine learning (ML) techniques to address it; in this work, we focus on \emph{Gaussian process} (GP) regression \cite{rasmussen2006gaussian}, including both standard GP regression and a version of GP regression designed to approximate monotonic functions \cite{riihimaki2010gaussian}. GP regression is a widely used non-parametric method for learning an approximate model from data and it is well suited to interpolation in low dimensions (small $\chi$) and with data from few circuits\cite{rasmussen2006gaussian}. We anticipate this regime to be the most likely in which featuremetric benchmarking is applied. GP regression is therefore a promising approach to analyzing the featuremetric benchmarking data, and, in the setting of volumetric benchmarking, we find that GPs can reproduce a volumetric benchmarking plot with high accuracy using much less data.

The rest of this paper is structured as follows. Section~\ref{sec:background} reviews \emph{capability functions}  \cite{Hothem2024-rc, Hothem2023-pc, Hothem2024-fc}, with which we formalize our problem, and volumetric benchmarking. Section~\ref{sec:fmb} presents featuremetric benchmarking, i.e., our general framework for capability benchmarking based on circuit features. Section~\ref{sec:examples} presents examples of featuremetric benchmarks, and the IBM Q and IonQ datasets used in the follow section. Section~\ref{sec:analysis} discusses the problem of how to analyze featuremetric benchmarking data, and presents our GP regression methods for doing so. We discuss avenues for improving the framework in Section~\ref{sec:discussion}.

\section{Background}\label{sec:background}
\subsection{Capability learning}\label{sec:caplearn}
The purpose of many quantum computer benchmarks, including our framework and volumetric benchmarking, can be elegantly formalized in terms of \emph{quantum capability learning} \cite{Hothem2024-rc, Hothem2023-pc, Hothem2024-fc}. Throughout this work, we assume that a quantum computer is \emph{stable}, i.e., we ignore the presence of slow drift and other time-varying errors that are common in contemporary quantum computers \cite{Proctor2020-iz}. This is convenient because it implies that performance on a particular quantum circuit $c$ is a constant, well-defined property of that system. Instability causes both foundational and practical problems in benchmarking that are considered elsewhere \cite{Hothem2024-rc, Proctor2021-wt}.

The foundational objects that define quantum capability learning are:
\begin{enumerate}
\item A circuit set $\mathbb{C}$.
\item A performance metric $s : \mathbb{C} \to \mathbb{R}$, which we call a \emph{capability function}.
\end{enumerate}
The circuit set $\mathbb{C}$ contains all the circuits whose performance (on some real quantum computing system) we are interested in. $\mathbb{C}$ could be the set of all circuits that are implementable using the native gates of a particular system, or the set of all Clifford circuits, or a set containing circuits that implement a particular algorithm. Circuits can be defined at many levels of abstraction \cite{Proctor2024-av, Blume-Kohout2020-de}, e.g., circuits can be defined in terms of high-level many-qubit operations (like an $n$-qubit quantum Fourier transform) or the low-level gates for a particular system, and different amounts and kinds of compilation can be permitted before executing a circuit. Featuremetric benchmarking is intended to be applicable to circuits defined at all levels of abstraction and with all levels of compilation freedom, and so we will only discuss the circuit abstraction level further when relevant, such as within our examples.

The capability function $s(c)$ quantifies how well a circuit $c \in \mathbb{C}$ was or could be executed on a particular system.  Examples for $s$ that are commonly used include success probability \cite{Hothem2024-rc, Proctor2021-wt} for definite-outcome circuits, like those of randomized benchmarking, process fidelity \cite{Hothem2024-rc, Hothem2024-fc, Proctor2022-zs}, and classical fidelity (also known as Hellinger fidelity) \cite{Lubinski2023-zy}.

Our featuremetric benchmarking framework is designed for probing $s(c)$ with arbitrary choices for $\mathbb{C}$ and $s$, i.e., it can be stated for general $s$ and $\mathbb{C}$. However, it is based on estimating $s(c_i)$ for each circuit $c_i$ in some (small) set of circuits $\{c_i\}$, and so it is necessary to be able to \emph{efficiently} estimate $s(c_i)$, i.e., the total resources needed to estimate $s(c)$ should not be so large as to be impractical. This is not the case for all choices of $s$ and $\mathbb{C}$, e.g., estimating the classical fidelity of a circuit $c$ general requires classically simulating $c$, which is infeasible for general circuits, as well as many repetitions of $c$. In our examples of featuremetric benchmarking, we use the following two choices for $s$ and $\mathbb{C}$.

\subsubsection{Success probability and mirror circuits}\label{sec:mirror-circuits}
We consider capability learning with a circuit set $\mathbb{C}$ containing only \emph{mirror circuits}  \cite{Proctor2021-wt, Proctor2022-yl,Hines2023-vq, Mayer2021-vl} and the capability function $s$ given by success probability. Mirror circuits are a kind of definite-outcome circuit. That is, each circuit $c \in \mathbb{C}$ is associated with a single bit string $x_c$ such that every execution of $c$ will return $x_c$ if $c$ is run without errors. One natural capability function for definite-outcome circuits (which is ill-defined for general circuits) is the success probability. This is defined by
\begin{equation}
    s(c) = \textrm{P}_c(x_c), \label{eq:sp}
\end{equation}
where $\textrm{P}_c(x)$ denotes the probability that circuit $c$ returns the bit string $x$. This is easily estimated from data obtained from finitely many executions of $c$, with the natural estimator being
\begin{equation}
    \hat{s}(c) = \frac{\# \textrm{ of times } x_c \textrm{ is output by } c}{N}, \label{eq:sp-estimate}
\end{equation}
where $N$ is the number of executions of $c$. This is the definition for the capability function, and the method for estimating its value, used in our experiments on IBM Q. The particular class of mirror circuits we use is discussed in Section~\ref{sec:examples}.

\subsubsection{Process fidelity and general Clifford circuits}\label{sec:general-circuits}
We will also consider capability learning with a circuit set containing circuits that contain only Clifford gates but are otherwise general, together wtih the capability function given by process fidelity. Process fidelity is a widely-used metric for quantifying the error in the quantum evolution implemented by a quantum circuit (or gate) \cite{Hashim2024-om}, and, unlike success probability, it is well-defined for general circuits. Process fidelity is also known as entanglement fidelity, and, for an $n$-qubit circuit $c$, it is defined by
\begin{equation}
    F(c)  =  \langle \varphi | (\mathbb{I} \otimes \mathcal{E}(c))[|\varphi \rangle \langle \varphi |]|\varphi \rangle.
\end{equation}
Here $\mathcal{E}(c)=\Lambda(c)\mathcal{U}^{\dagger}(c)$ is the circuit $c$'s error map, where $\mathcal{U}(c)$ is the superoperator representation of the unitary that $c$ ideally implements and $\Lambda(c)$ is the superoperator representing the imperfect evolution actually implemented, and $\varphi$ is any maximally entangled state of $2n$ qubits \cite{Hashim2024-om}. Note that $F$ is equal to a linear rescaling of average gate set fidelity $\bar{F}$, which is another widely-used metric for error \cite{Hashim2024-om}. Therefore, learning an approximation for $F(c)$ is equivalent to learning an approximation for $\bar{F}(c)$.

The process fidelity with which a circuit $c$ is executed cannot be reliably estimated by simply running $c$. Instead, we require a more complex procedure for estimating $s(c)$, in which $c$ is embedded within other circuits. There is a technique for efficiently estimating $s(c_i)$ for (almost) arbitrary circuits $c_i$: mirror circuit fidelity estimation (MCFE) \cite{Proctor2022-zs}. However, in this work, we choose to focus on \emph{Clifford} circuits in our experiments for learning $F(c)$, on an IonQ system. For Clifford circuits, there is an alternative and even more efficient method for estimating $F(c)$: An adaptation to direct fidelity estimation (DFE) \cite{Flammia2011-qj,Moussa2012-rq} introduced in Ref.~\cite{Seth2025-zz} that we will call SPAM-error-robust DFE (SR-DFE). SR-DFE requires running fewer circuits than MCFE, and it is described in Appendix~\ref{app:dfe}.

\subsection{Volumetric benchmarking}\label{sec:vb}
Volumetric benchmarking~\cite{Blume-Kohout2020-de, Proctor2021-wt} is a framework for benchmarking quantum computers, rather than a specific benchmark. It is a scheme for constructing benchmarking circuits and summarizing the data obtained. A volumetric benchmark consists of running circuits of varied widths and depths, and then plotting observed performance in the width $\times$ depth plane---which we will call a \emph{volumetric benchmarking plot}---as demonstrated in Figure~\ref{fig:ibm_montreal_vb}a.

A specific volumetric benchmark is defined by a circuit family $\mathbb{C}_{w,d} \subset \mathbb{C}$, indexed by circuit width ($w$, i.e., the number of qubits) and depth ($d$), and a capability function $s(c)$. Examples of circuit families for which a volumetric benchmark can be defined include the quantum volume circuits \cite{Cross2019-ku, Hines2024-ae}, randomized mirror circuits \cite{Proctor2021-wt, Proctor2022-yl, Hines2023-vq} (which we used in the experiments summarized in Figure~\ref{fig:ibm_montreal_vb}), or the circuits from many algorithms, as in the benchmarking suite developed by members of the QED-C \cite{Lubinski2023-zy}. Applying a volumetric benchmark to a quantum computer consists of picking a range of circuit shapes $(w,d)$, selecting specific circuits for each shape (e.g., via sampling from a distribution) from $\mathbb{C}_{w,d}$, and running them (or closely related circuits) to estimate $s(c)$ for each circuit $c$.

Volumetric benchmarking is simple and intuitive, but it has some important limitations that stem from its \emph{descriptive} rather than \emph{predictive} quality. The central tenant of volumetric benchmarking is to plot circuit performance versus circuit shape. However, not all circuits with the same shape within a family necessary have the same performance. Figure~\ref{fig:ibm_montreal_vb}a shows the \emph{mean} success probability of 20 different randomized mirror circuits of each shape, but there is significant variation in the success probabilities for circuits of the same shape, as shown in Fig.~\ref{fig:ibm_montreal_vb}b. This limits the predictive power of a volumetric benchmarking plot, as, given some circuit of interest $c \in \mathbb{C}$, the mean success probability shown in Fig.~\ref{fig:ibm_montreal_vb} at $c$'s width and depth will generally inaccurately predict $c$'s success probability.

Circuit width and depth are just two possible \emph{features} of a circuit, and additional (or different) features might result in lower variance in the performance of circuits with the same values for those features, and therefore enable more accurate predictions of performance. This is the central idea that motivates the featuremetric benchmarking framework.

\section{Featuremetric benchmarking} \label{sec:fmb}
We now introduce our featuremetric benchmarking framework. Featuremetric benchmarking consists of 
\begin{enumerate}
\item[(i)] picking a small set of circuit features of interest, e.g., that we anticipate will be approximately predictive of $s(c)$, 
\item[(ii)] varying the values of those circuit features and sampling circuits for them, and 
\item[(iii)] running those circuits to observe how $s(c)$ depends on our features.
\end{enumerate}
In this section, we introduce each of the concepts needed to describe featuremetric benchmarking more precisely (Sections~\ref{sec:features}-\ref{sec:circuit-sampling}), leading to a more technical statement of the above procedure (Section~\ref{sec:fbm-method}). Readers who are satisfied with the above description of featuremetric benchmarking may wish to skip to Section~\ref{sec:examples}.

\subsection{Circuit features}\label{sec:features}
The foundational concept in featuremetric benchmarking is the circuit feature, and so we now define this concept. A \emph{circuit feature} is simply a function $f:\mathbb{C} \to \mathbb{R}$. It takes a circuit and computes a number $f(c)$ that partially describes that circuit. Any such function defines a valid circuit feature, and a very large number of features will completely describe each $c \in \mathbb{C}$. In our context, we are interested in a \emph{small} set of circuit features that are approximately predictive of $s(c)$, i.e., if we know only the values of those features for a circuit $c$ we can predict $s(c)$ with reasonable accuracy. This means that we are interested in circuit features that are predictive of the amount of error that occurs when $c$ is executed.

It is not currently clear whether it is possible to reliably predict $s(c)$ from a small set of features, or what good features are \cite{Proctor2024-av}. In principle, the most predictive features will depend on the quantum computer being modeled, but it is plausible that a small, fixed set of features will be sufficient to predict $s(c)$ in many systems. Examples of interesting circuit features that tend to be somewhat predictive of $s(c)$ include:
\begin{enumerate}
    \item The depth $d(c)$ of $c$, which is the number of layers of gates in $c$ (and which is one of the two features used in volumetric benchmarking)
    \item The width $w(c)$ of $c$, which is the number of qubits on which $c$ acts (and which is the other feature used in volumetric benchmarking).
    \item The number of two-qubit gates $N_{\textrm{2Q}}(c)$ in $c$, or their density $\xi_{\textrm{2Q}}(c) = 2 N_{\textrm{2Q}}(c)/(w(c)d(c))$. Density is often more convenient than $N_{\textrm{2Q}}(c)$ because interesting circuit classes can be defined in which $\xi_{\textrm{2Q}}(c)$, width and depth can be varied independently (whereas for interesting circuit classes $N_{\textrm{2Q}}(c)$ will typically grow with width and depth).
    \item The number of single-qubit gates $N_{\textrm{1Q}}(c)$ in $c$, or their density $\xi_{\textrm{1Q}}(c) = N_{\textrm{1Q}}(c)/(w(c)d(c))$.
    \item The number of mid-circuit measurements $N_{\textrm{MCM}}(c)$ in $c$, or their density $\xi_{\textrm{MCM}}(c) = n_{\textrm{MCM}}N_{\textrm{MCM}}(c)/(w(c)d(c))$ where $n_{\textrm{MCM}}$ is the number of qubits on which each mid-circuit measurement acts.
    \item Whether or not qubit $q$ is acted on by $c$, which can be formalized in the function $A_q(c) \in \{0,1\}$. For a system with $n$ qubits, denoted by $q=1,\dots,n$, the $n$ features $A_1(c)$, $A_2(c)$, $\dots$, $A_n (c)$ together specify which qubits $c$ acts on.
\end{enumerate}
Other examples of circuit features are given by Tomesh \emph{et al.}~\cite{Tomesh2022-nu}.

Precisely defining a particular circuit feature is sometimes challenging, e.g., seemingly intuitive circuit features such as circuit depth can be defined in a variety of reasonable ways (see Refs.~\cite{Blume-Kohout2020-de, Proctor2021-wt} for detailed discussions of this). Similarly, a feature may only be well-defined for circuits defined at a particular level of abstraction. For example, qubit labels in a high-level circuit are typically arbitrary place-holders as that circuit must be compiled into a particularly systems native gates and assigned to  physical qubits (sometimes called ``qubit routing''), so the $A_i(c)$ features are arguably ill-defined for high-level circuits. 

\subsection{Circuit pseudo-features}\label{sec:pseudo-features}
There is another, complementary notion for a circuit feature---a \emph{circuit pseudo-feature}---that we also find to be useful in practice. A benchmark is often specified using a (classical) algorithm $\mathcal{A}$ for constructing or sampling circuits and, in general, this algorithm is parameterized by some variables $\vec{w}$. For example, we might be interested in creating a featuremetric benchmark that studies $s(c)$ for a family of quantum approximate optimization algorithm (QAOA) circuits for the graph \texttt{MaxCut} problem. The classical algorithm for constructing these circuits will be parameterized by a variety of parameters within the QAOA algorithm itself as well as various implementation details. Parameters include the size of the problem graph and the number of QAOA layers (typically denoted $p$) \cite{Lubinski2023-mr}. Neither the problem graph or $p$ are strictly circuit features, as they are not functions of the produced circuit $c$. In our examples in Section~\ref{sec:examples}, we will encounter another example of a pseudo-feature: the mean $\xi_{\textrm{2Q}}$ for a circuit ensemble.

The parameters $\vec{w}$ of a circuit-generating algorithm $\mathcal{A}$ are not necessarily true circuit features, because from a circuit $c$ produced by $\mathcal{A}$ there might not be a function $f_i$ such that $f_i(c) = w_i$, i.e., we cannot uniquely identify each $w_i$ given $c$. $s(c)$ can only depend on true features of $c$---i.e., it does not matter how $c$ was created---but circuit pseudo-features are often closely related to true circuit features (e.g., QAOA's $p$ is closely related to circuit depth). Furthermore, it is common in benchmarking to ask how performance varies with quantities such as algorithmic depth \cite{Lubinski2023-zy, Lubinski2023-mr}, or with Trotter step size for Hamiltonian simulation circuits \cite{Chatterjee2024-py}. It is therefore useful to include pseudo-features within the featuremetric benchmarking framework. Throughout the rest of this paper, we distinguish between true circuit features and circuit pseudo-features only when this distinction is critical.

\subsection{Feature vectors}\label{sec:feature-vectors}
In featuremetric benchmarking we vary the values of some set of features, and we denote our chosen features by $f_1$, $f_2$, $\dots$, $f_{\chi}$ (so $\chi$ is the number of features) and we arrange them into a vector
\begin{equation}
\vec{f} = (f_1,f_2, \dots, f_{\chi}).
\end{equation}
Each feature $f_i$ can take on some set of values $\mathbb{I}_i \subseteq \mathbb{R}$ (i.e., $\mathbb{I}_i$ is $f_i$'s image). For example, the circuit width $w(c)$ must be a positive integer, and it is often interesting to consider a circuit set $\mathbb{C}$ such that the circuit width $w(c)$ takes all values in $\{1,2,\dots,n\}$ where $n$ is the number of qubits available.

In featuremetric benchmarking, we select $M$ different feature vectors $\vec{v}_1$, $\vec{v}_2$, $\dots$, $\vec{v}_M$ from $\mathbb{I} = \mathbb{I}_1 \times \mathbb{I}_2 \cdots \mathbb{I}_{\chi}$, and run circuits based on those feature vectors. Below we discuss how we pick circuits to run given a feature vector $\vec{v}$, but first we briefly discuss how we select the $M$ vectors $\vec{v}_i$. In volumetric benchmarking, $\chi=2$ and the feature vectors are systematically varied through some 2-dimensional grid of values (see Figure~\ref{fig:ibm_montreal_vb}). For a grid with $k$ divisions, the number of required feature vectors $M$ scales as $M = \mathcal{O}(k^{\chi})$ and so this is not feasible except for very small $\chi$. However, many alternative approaches for selecting our feature vectors exist, such as sampling from a distribution over $\mathbb{I}$. The merits of a particular strategy are intrinsically linked to what exactly we aim to extract from the data. We discuss some example strategies in Section~\ref{sec:analysis}.

\subsection{Sampling circuits from feature vectors}\label{sec:circuit-sampling}
Given a feature vector $\vec{v}$, we must select circuits to run. In featuremetric benchmarking, we sample circuits from a user-chosen distribution $P_{\vec{v}}$ over $\mathbb{C}$ that is parameterized by $\vec{v}$ (or, in the case of pseudo-features, $P_{\vec{v}}$ is defined over the set of circuits produced by $\mathcal{A}$ with those pseudo-features). We say that $P_{\vec{v}}$ is \emph{user-chosen} because, just like volumetric benchmarking, featuremetric benchmarking is designed to encompass benchmarks based on many different kinds of circuits---e.g., randomized mirror circuits \cite{Proctor2021-wt, Proctor2022-yl,Hines2023-vq, Mayer2021-vl}, periodic mirror circuits \cite{Proctor2021-wt}, quantum-volume-like circuits \cite{Cross2019-ku, Hines2024-ae, Amico2023-ze}, QAOA circuits, and brickwork circuits like those of cross-entropy benchmarking \cite{Boixo2018-kp}. The circuit type is encoded into $P_{\vec{v}}$ (as well as $\mathbb{C}$). But $P_{\vec{v}}$ cannot be arbitrarily chosen, because the point of featuremetric benchmarking is that we are aiming to explore how our circuit family's performance varies with our feature vector $\vec{v}$.

The simplest and most intuitive property to require of $P_{\vec{v}}$ is that any circuit sampled from $P_{\vec{v}}$ has the value $\vec{v}$ for the feature $\vec{f}$, i.e., if $c$ is sampled from $P_{\vec{v}}$ then
\begin{equation}
    \vec{f}(c) = \vec{v},\label{eq:condition-1}
\end{equation}
with certainty, where
\begin{equation}
\vec{f}(c) \equiv (f_1(c),f_2(c), \dots, f_{\chi}(c)) \in \mathbb{I}.
\end{equation}
However, this requirement turns out to be too strong in practice, because it is easy to come up with interesting choices for $\vec{f}$ where we cannot find any circuit that satisfies Eq.~\eqref{eq:condition-1} for some feature vectors $\vec{v}$. Consider the feature set of width $w(c)$, depth $d(c)$, and two-qubit gate density $\xi_{\textrm{2Q}}(c)$. For small widths and depths, $\xi_{\textrm{2Q}}(c)$ is restricted to a small set of values, e.g., for $w(c)=2$ and $d(c)=2$ the only possible values for $\xi_{\textrm{2Q}}(c)$ are 1, 0.5, and 0. This means that we cannot independently vary our three features and exactly satisfy Eq.~\eqref{eq:condition-1}. For large $w$ and $d$, $\xi_{\textrm{2Q}}$ is approximately continuous and can be varied independently of $wd$, but this is never exactly true. Additionally, we are sometimes interested in relatively small $wd$, so we instead specify a condition on $P_{\vec{v}}$ that we can satisfy exactly.

We require that $P_{\vec{v}}$ is a distribution for which the \emph{expected value} of $\vec{f}$ for circuits sampled from $P_{\vec{v}}$ is $\vec{v}$. That is, if $C$ is distributed according to $P_{\vec{v}}$ (so $C$ is a circuit-valued random variable) then
\begin{equation}
    \mathbb{E}(f_i(C)) = v_i.
\end{equation}
Where possible, however, it is typically more useful if we satisfy Eq.~\eqref{eq:condition-1}. Or, stated more generally, it is preferable to keep the variance of each $f_i(C)$ small. This is because we are typically interested in whether the features $\vec{f}$ are sufficient to predict $s(c)$ for a circuit $c$, and we can only test this directly if we run multiple circuits with the same $\vec{v}$ (or very nearly the same $\vec{v}$).

\begin{figure}[t!]
    \centering  
    \includegraphics[width=8.cm]{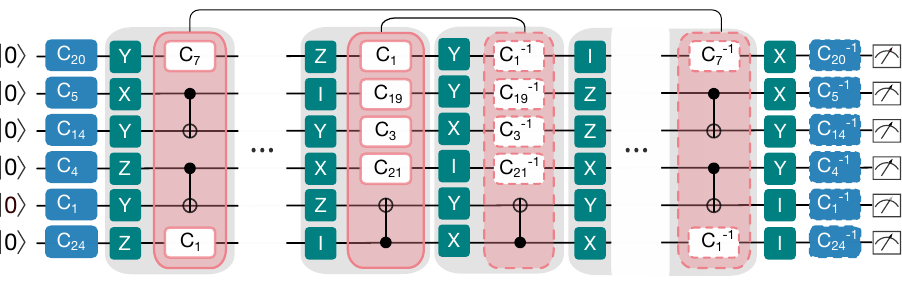}
    \caption{\textbf{Randomized mirror circuits}. A diagram of the randomized mirror circuits used in two of our example featuremetric benchmarks, both of which we ran on IBM Q systems. These particular randomized mirror circuits contain only Clifford gates ($C_i$, with $i=0,1,\dots,23$, denote the 24 single-qubit Clifford gates). They have a variable width, depth, and mean density of two-qubit gates.}
    \label{fig:rmcs}
\end{figure}

\begin{figure*}[t!]
    \centering  
    \includegraphics[width=18cm]{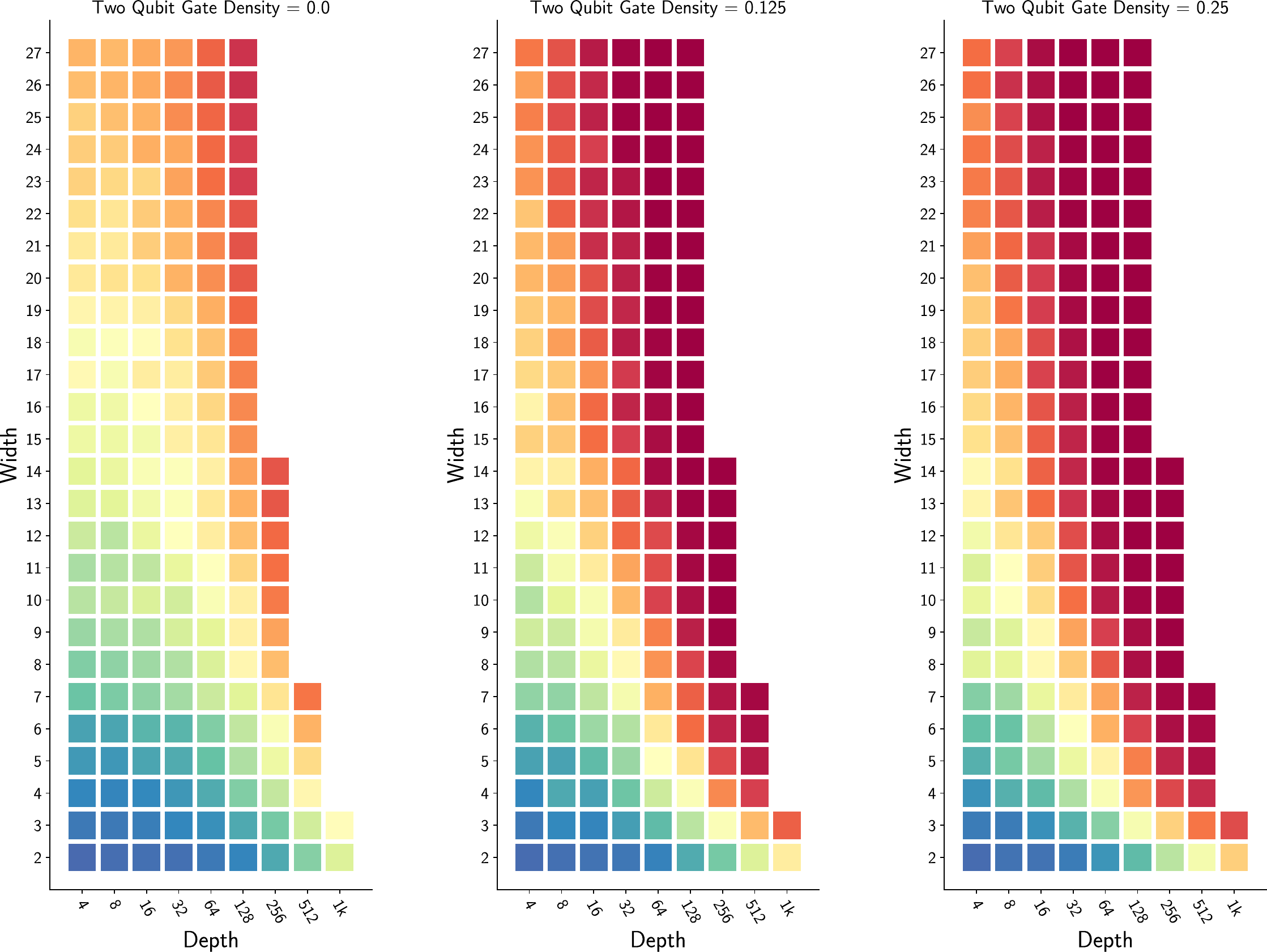}
    \caption{\textbf{A simple featuremetric benchmark summarized in volumetric benchmarking plots (\texttt{ibmq\_algiers})}. An example of a simple featuremetric benchmark, with three features, in which the data can be summarized with three volumetric benchmarking plots, which we ran on \texttt{ibmq\_algiers}. This featuremetric benchmark consists of varying three circuit features: circuit width ($w$), circuit depth ($d$), and two-qubit gate density ($\xi_{\textrm{2Q}}$), and we systematically varied the features $(w, d, \xi_{\textrm{2Q}})$ over a three-dimensional grid of values $\{2,3,\dots,27\} \times \{2^k\}_{i =2,3,\dots 10} \times \{0, 1/8, 1/4\}$ (except that feature vectors with large depths and widths were discarded, with the particular values discarded implied by the missing data in the plot). Because $\xi_{\textrm{2Q}}$ took only three discrete values (0, $1/8$ and $1/4$), and we systematically varied both circuit width and depth, we can represent the results in three volumetric benchmarking plots---one for each value of $\xi_{\textrm{2Q}}$---as shown here. The circuits used were randomized mirror circuits, and, as in the volumetric benchmarking results for \texttt{ibmq\_montreal} shown in Figure~\ref{fig:ibm_montreal_vb}, we plot the mean success probability of all $K=10$ randomized mirror circuits run at each feature value. We observe substantial changes in the circuits' success probabilities as we vary $\chi_{\textrm{2Q}}$. For example, at $(w,d)=(14,64)$ the mean success probabilities for $\chi_{\textrm{2Q}}=0$, $1/8$, and $1/4$ are $(51\pm 2)\% $, $(19 \pm 4)\%$, and $(7\pm 2)\%$, respectively, where, here and throughout, error bars are the standard error calculated using a bootstrap. This implies that our choice to vary two-qubit gate density, in addition to circuit width and depth, will increase the predictive accuracy of a model for circuit performance based around interpolating these results. }
    \label{fig:ibm_algiers_vb}
\end{figure*}

\subsection{The featuremetric benchmarking method}\label{sec:fbm-method}
We now have introduced all of the components needed to precisely state the featuremetric benchmarking methodology. A featuremetric benchmark is defined by
\begin{enumerate}
    \item A set of $\chi$ features $\vec{f}=(f_1,f_2,\dots, f_{\chi})$.
    \item A procedure for selecting $M$ feature vectors $\vec{v}_1$ , $\vec{v}_2$, $\dots$, $\vec{v}_M$.
    \item A probability distribution $P_{\vec{v}}$ over $\mathbb{C}$, from which we sample $K$ circuits for each $\vec{v}_i$.
\end{enumerate}
The circuit sampling process described above creates a set of $MK$ circuits, $\{c_{i,j}\}$, where $c_{i,j}$ is the $j$\textsuperscript{th} circuit sampled from the distribution $P_{\vec{v}_i}$, i.e., with the $i$\textsuperscript{th} value for the feature vector. 

A featuremetric benchmarking experiment consists of running circuits to estimate $s(c_{i,j})$ for each of our circuits $c_{i,j}$. This can simply mean executing each $c_{i,j}$ circuit $N \gg 1$ times, as is the case for the two IBM Q data sets that we present (for those data sets $s$ is success probability and it is estimated via Eq.~\eqref{eq:sp-estimate}). However, more generally, $s(c_{i,j})$ might be measured indirectly, by running some other closely-related circuits---as is the case if $s$ is process fidelity and $s$ is estimated via mirror circuit fidelity estimation \cite{Proctor2022-zs}, Cliffordization \cite{Ferracin2021-vh, Seth2025-zz}, or direct fidelity estimation \cite{Flammia2011-qj, Moussa2012-rq}.

\section{Example featuremetric benchmarks}\label{sec:examples}
We now demonstrate featuremetric benchmarking with three experiments that we ran on IBM Q and IonQ cloud-access quantum computers.

\subsection{Featuremetric benchmarking with two features}\label{sec:experiment-montreal}
Featuremetric benchmarking is a generalization of volumetric benchmarking, and our first example of a featuremetric benchmark is also a volumetric benchmark. We include this example because our techniques for analyzing featuremetric benchmarking data (Section~\ref{sec:analysis}) are also of interest for the special case of volumetric benchmarking (and they are particular easy to visualize in this case). Our example of volumetric benchmarking consists of running \emph{randomized mirror circuits} \cite{Proctor2021-wt, Proctor2022-yl,Hines2023-vq, Mayer2021-vl}, and we ran this volumetric benchmark on \texttt{ibmq\_montreal}, a cloud-accessible 27-qubit transmon-based quantum computer (this experiment was performed in July 2021, with the data also used in Refs.~\cite{Hothem2023-pc,Hashim2024-om}). We now describe this volumetric benchmark in detail, as it is illustrative of our framework and we will build on it in our later, more complex, examples. IBM Q's calibration data for \texttt{ibmq\_montreal}, at the time of this experiment, are provided in Appendix~\ref{app:caldata}.

Randomized mirror circuits are mirror circuits with random layers \cite{Proctor2022-yl, Proctor2021-wt, Hines2023-vq} and various easily-controllable circuit features that include their width and depth. There are many possible kinds of randomized mirror circuits, and so we describe the particular family we use (which are the same as some of those used in Ref.~\cite{Proctor2022-yl, Proctor2021-wt}). These particular randomized mirror circuits contain only Clifford gates and, for a set of qubits $\mathbb{Q}$, are defined by a method $\mathcal{S}_{\mathbb{Q}}$ for sampling a random layer of gates to apply to $\mathbb{Q}$. Given such a sampling method $\mathcal{S}_{\mathbb{Q}}$, a \emph{benchmark depth} $d$ randomized mirror circuit consists of (i) $\nicefrac{d}{4}$ layers of gates independently sampled using $\mathcal{S}_{\mathbb{Q}}$, followed by those layers in reverse with each gate replaced with its inverse, (ii) layers of uniformly random Pauli gates interleaved in between all $\nicefrac{d}{2}$ layers, and (iii) an initial layer of uniformly-random single-qubit Clifford gates, and its inverse at the end of the circuit \footnote{These layers are a standard part of mirror circuits that are particularly important when using them in a RB method for measuring layer error rates \cite{Proctor2022-yl, Hines2023-vq}, but they are not important for our purposes here.}. The structure of our randomized mirror circuits is shown in Fig.~\ref{fig:rmcs}.

In our experiments, our method for sampling layers ($\mathcal{S}_{\mathbb{Q}}$) is as follows. A layer of gates is constructed so that, on average, it contains a density of two-qubit gates of $2\xi$. Two-qubit gates are only applied between qubits in $\mathbb{Q}$ that are connected \footnote{The particular method we use for randomly selecting which two-qubit gates to include is the \emph{edgegrab} algorithm described in the Supplemental Material of Ref.~\cite{Proctor2021-wt}.}---the circuits are specifically designed for the system being benchmarked. All qubits not acted on by a two-qubit gate are assigned an independent and uniformly random single-qubit gate from the set of all 24 single-qubit Clifford gates. This means that a randomized mirror circuit sampled in this way has an average two-qubit gate density of $\xi$ \footnote{Where density is defined with respect to benchmark depth $d$, but note that the circuit's depth could also reasonable be defined as $d + 3$.}. $\xi$ is a parameter of the circuit sampling (it is a circuit \emph{psuedo-feature}) and so is $\mathbb{Q}$ (as described in Section~\ref{sec:features}, this can be represented by $n$ circuit features). We therefore denote the sampling distribution by $P_{\textsc{rmc}, (w,d, \xi_{\textrm{2Q}}, \mathbb{Q})}$. In these experiments we vary only two circuit features---width and depth---so we fix $\xi$ to $\xi = \nicefrac{1}{4}$ and we pick the qubit set $\mathbb{Q}$ to be the first $w$ qubits of the system $\mathbb{Q}_{1:w} = \{q_1, \dots, q_w\}$ (according to IBM's labeling of their qubits). That is, we sample circuits from $P_{\textsc{rmc}, (w,d, 0.25, \mathbb{Q}_{1:w})}$.

We vary the features $(w,d)$ over a two-dimensional grid of values:
\begin{equation}
    \vec{v} \in \{2,3,\dots,27\} \times \{2^k\}_{i =2,3,\dots 12}.
\end{equation}
We include all such values for $\vec{v}$ \emph{except} that we exclude some of the larger depths for larger widths (Figure~\ref{fig:ibm_montreal_vb} implies the widths and depths that we included). At each feature vector value, $(w,d)$, we sample $K=20$ randomized mirror circuits to run on the first $w$ qubits of \texttt{ibmq\_montreal}, using the procedure described above. Each circuit is repeated $N=1024$ times to estimate its success probability (using Eq.~\eqref{eq:sp-estimate}). The resulting data is summarized in the volumetric benchmarking plot of Figure~\ref{fig:ibm_montreal_vb}.

\subsection{Three features: width, depth, and two-qubit gate density}\label{sec:experiment-algiers}
We now present an example of a featuremetric benchmark with three circuit features, which is a simple generalization of the volumetric benchmark described above. In many contemporary quantum computing systems, two-qubit gates have much higher error rates than single-qubit gates, and in those systems the density of two-qubit gates in a circuit $\xi_{\textrm{2Q}}$ will have a substantial effect on that circuit's error rate. Therefore, $\xi_{\textrm{2Q}}$ is a natural candidate for a circuit feature with which to define a featuremetric benchmark, in addition to circuit depth and width. We design a featuremetric benchmark based on randomized mirror circuits using the three-dimensional feature vector $(w, d, \xi_{\textrm{2Q}})$, to run on \texttt{ibmq\_algiers}, which is a 27-qubit system similar to \texttt{ibmq\_montreal}.

We vary the features $(w, d, \xi_{\textrm{2Q}})$ over a three-dimensional grid of values: 
\begin{equation}
    \vec{v} \in \{2,3,\dots,27\} \times \{2^k\}_{i =2,3,\dots 10} \times \{0, 1/8, 1/4\}.
\end{equation}
We include all such values for $\vec{v}$ \emph{except} that we exclude some of the larger depths for larger widths (Figure~\ref{fig:ibm_algiers_vb} shows the widths and depths that we included). At each feature vector value, we sample randomized mirror circuits with those circuit features. These randomized mirror circuits had the same structure as those in Section~\ref{sec:experiment-montreal}, i.e., they are defined by the same distribution over circuits $P_{(w,d, \xi_{\textrm{2Q}}), \mathbb{W}_{1:w}}$ but now with $\xi_{\textrm{2Q}}$ varied. We ran this benchmark on \texttt{ibmq\_algiers} with $K=10$ randomized mirror circuits at each feature value. Each circuit was executed 1024 times, and, as with \texttt{ibmq\_montreal}, width $w$ circuits were run on the first $w$ qubits of \texttt{ibmq\_algiers}, using IBM's labeling of this system's qubits. IBM Q's calibration data for \texttt{ibmq\_algiers}, at the time of this experiment, are provided in Appendix~\ref{app:caldata}.

\begin{figure*}[t!]
    \centering  
    \includegraphics[width=18cm]{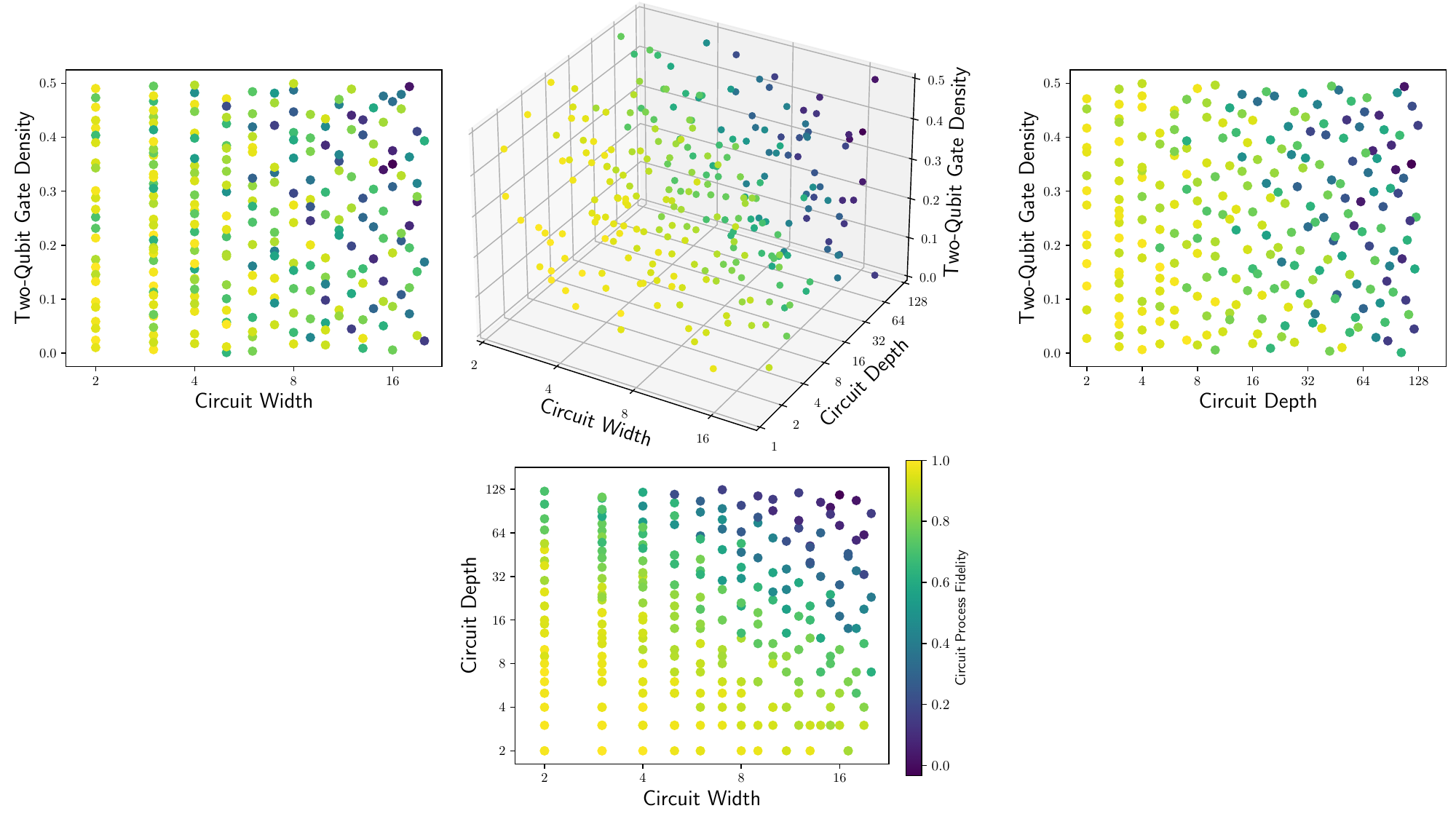}
    \caption{\textbf{A three-dimensional featuremetric benchmark (\texttt{Forte1})}. The results of a three-dimensional featuremetric benchmark run on 20 qubits of IonQ's \texttt{Forte1} cloud-access system. In this benchmark, we measured the process fidelities $F$ of random circuits versus three circuit features: width, depth, and two-qubit gate density. In the central panel, we show the mean estimated $F$ at each feature vector value $\vec{v}$ (we selected and measured the process fidelities of 30 circuits at each $\vec{v}$) versus the three feature values. We also show three different 2-dimensional projections of the data, each consisting of discarding one of the three features. In this benchmark, feature vectors were selected quasirandomly, using a Sobol sequence, so as to more uniformly ``fill up'' the 3-dimensional feature space than is typical with (pseudo)randomly sampled feature vectors.}
    \label{fig:ionq_forte_data}
\end{figure*}

In this featuremetric benchmark we are (almost) exhaustively varying $\chi$ features over a $\chi$-dimensional grid of $\alpha_i$ different values for feature $i$, which means running circuits with $\prod_{i=1}^{\chi} \alpha_i$ different feature values. The number of circuits to run therefore grows very quickly with $\chi$, and even in this example---with only $\chi =3$ and only three values (i.e., $\alpha_i = 3$) for one of our three features---the number of feature vector values is fairly large. Specifically, in our experiment we ran circuits at 531 different feature vector values, with $K=10$ circuits at each feature value, for a total of 5310 circuits. Below, we introduce an approach to designing featuremetric benchmarking experiments that enables running circuits with fewer feature values. However, one advantage of this experiment design is that we can easily visualize the data using three volumetric benchmarking plots---one for each value of $\xi_{\textrm{2Q}}$. These plots are shown in Figure~\ref{fig:ibm_algiers_vb}, which show the mean success probability for the $K=10$ randomized mirror circuits run at each feature vector value.

We observe that the success probabilities of equal-shape circuits vary significantly as $\xi_{\textrm{2Q}}$ varies. For example, at $(w,d)=(14,64)$ the mean success probabilities for $\chi_{\textrm{2Q}}=0$, $1/8$, and $1/4$ are $(51\pm 2)\% $, $(19 \pm 4)\%$, and $(7\pm 2)\%$, respectively. Here and throughout, error bars are the standard error (1$\sigma$) calculated using a bootstrap. A dependency on $\xi_{\textrm{2Q}}$ is unsurprising, due to the significant difference between the error rates of one-qubit and two-qubit gates in contemporary IBM Q systems, but quantifying this dependency was enabled by the framework of featuremetric benchmarking, whereas it is ignored by volumetric benchmarks. Varying the two-qubit gate density, in addition to circuit width and depth, enables an increase in the predictive accuracy of a model for circuit performance based around interpolating these results.

In this benchmark, the two-qubit gate density $\xi_{\textrm{2Q}}$ is a circuit \emph{pseudo}-feature (see Section~\ref{sec:pseudo-features}). The two-qubit gate density parameterizes the distribution from which the circuits were sampled $P_{(w,d, \xi_{\textrm{2Q}},\mathbb{Q}_{1:w})}$, and the mean two-qubit gate density of circuits sampled from $P_{(w,d, \xi_{\textrm{2Q}},\mathbb{Q}_{1:w})}$ is $\xi_{\textrm{2Q}}$, but each circuit sampled from $P_{(w,d, \xi_{\textrm{2Q}},\mathbb{Q}_{1:w})}$ does not generally have a two-qubit gate density of $\xi_{\textrm{2Q}}$. In our next example, we present a featuremetric benchmark in which two-qubit gate density is a true circuit feature.

\subsection{Three features with random sampling}
Our final example featuremetric benchmark consists of varying the same three features---width, depth, and two-qubit gate density---but we choose the values of our features (quasi) \emph{randomly}. Furthermore, we use a circuit ensemble containing random Clifford circuits that do not have a definite outcome (so they do not have a well-defined success probability), with the circuit's \emph{process fidelity} as our capability function. The process fidelity of each selected circuit is estimated using SR-DFE (see Section~\ref{sec:general-circuits}).

The random circuits we use contain $d$ randomly sampled layers of one- and two-qubit Clifford gates gates. For a given feature vector $(w,d,\xi_{\textrm{2Q}})$, we sample a width-$w$ circuit with $d$ layers that deterministically contains $wd\xi_{\textrm{2Q}}/2$ two-qubit gates (except that we round this to an integer, if necessary), resulting in a two-qubit gate density of $\xi_{\textrm{2Q}}$. Therefore, in this featuremetric benchmark, two-qubit gate density is a true circuit feature instead of a pseudo-feature, i.e., all circuits sampled from this benchmark's $P_{(w,d, \xi_{\textrm{2Q}})}$ have the value of $(w,d,\xi_{\textrm{2Q}})$ for our three features, rather than just this value in expectation. To achieve this, our circuits cannot contain \emph{independently} sampled layers. Instead, we first randomly select locations within the circuit to include two-qubit gates, and populate all other circuit locations. This strategy results in circuits whose random layers have the same marginal distributions, but are not independent.

 Instead of exhaustively varying our feature vector as in our previous examples, we only sample values for $\vec{v}$ here. The simplest sampling strategy is (pseudo-)random sampling from $\mathbb{R}^{\chi}$ (or, more precisely, from $\vec{f}$'s range). However, a set of independent, randomly-sampled $\chi$-dimensional vectors are typically ``clustered'', i.e., they do not evenly fill the $\chi$-dimensional space from which they are sampled, with this clustering problem increasing as $\chi$ increases. So, instead, we select $\vec{v}$ \emph{quasirandomly}, i.e., we choose our $K$ feature vectors $\vec{v}_1$, $\vec{v}_2$, \dots, $\vec{v}_K$ from a \emph{quasirandom sequence} \cite{Kuipers2012-je}. Heuristically, a quasirandom sequence is a sequence that looks random except that the points are much more evenly distributed than in a typical random sequence. This strategy for selecting our feature vectors therefore enables a more sample-efficient experiment.

Because we expect circuit fidelity to decay approximately exponentially in both circuit width and depth, we more uniformly explore different circuit fidelity values if the logarithms of circuit width and depth are evenly spaced between some minimum and maximum. We therefore choose to select vectors $(w_{\log},d_{\log},\xi_{\textrm{2Q}})$ from a \emph{Sobol sequence} \cite{Sobol1967-lj} on $[w_{\min, \log},w_{\max, \log}] \times [d_{\min, \log},d_{\max, \log}] \times [\xi_{\min, \textrm{2Q}},\xi_{\max, \textrm{2Q}}]$ and set our feature vector to $(w ,d,\xi_{\textrm{2Q}}) = (2^{w_{\log}},2^{d_{\log}},\xi_{\textrm{2Q}})$ (with width and depth rounded to the nearest integer). A Sobol sequence is one of the many possible quasirandom sequences. The parameters $w_{\min, \log}$, $w_{\max, \log}$, $d_{\min, \log}$, $d_{\max, \log}$, $\xi_{\min, \textrm{2Q}}$, and $\xi_{\max, \textrm{2Q}}$ control the minimum and maximum possible value for each feature, and in the featuremetric benchmark that we ran we set them so that the minimum and maximum circuit widths were $2$ and $20$, respectively, the minimum and maximum circuit depths were 2 and 128, respectively, and the minimum and maximum two-qubit gate densities were 0 and $\nicefrac{1}{2}$, respectively. The quasirandom feature vectors used in our experiment are shown in Figure~\ref{fig:ionq_forte_data}.

We performed this benchmarking on IonQ’s \texttt{Forte1}, a cloud-accessible trapped-ion quantum processor with up to 36 qubits and coherence time ($T_2$) up to 1 second. The medians of its single-qubit gate error rates, two-qubit gate error rates, and SPAM error rates are 0.03\%, 0.75\%, and 0.43\%, respectively. Gate durations are 130~$\mu$s for single-qubit gates and 970~$\mu$s for two-qubit gates. We selected $K=30$ circuits to run at each of $256$ different feature vector values. Together with the SPAM error reference circuits used in SR-DFE (see Section~\ref{sec:general-circuits}), this resulted $2 \times 30 \times 256 = 15360$ circuits to execute. Each circuit was repeated 200 times, and each circuit's process fidelity was estimated using the analysis of SR-DFE.

Figure~\ref{fig:ionq_forte_data} shows the estimated mean process fidelities at each of the 256 different feature values, obtained in this experiment. In the main panel of Fig.~\ref{fig:ionq_forte_data}, we show the estimate circuit fidelities versus the circuit's features, in the 3-dimensional feature space. Figure~\ref{fig:ionq_forte_data} also shows the three projections of this data into 2-dimensions obtained by discarding one of the three features. We observe that if we discard any of the three circuit features we reduce our ability to predict a circuit's process fidelity: each 2-dimensional projection is not monotonic, i.e., the values of the process fidelities do not uniformly decrease with increasing feature values. This is most significant when we discard either circuit width or depth, but is also true for the case of discarding two-qubit gate density. The 3-dimensional data is also not statistically consistent with a monotonic decay in process fidelity with increase feature values, but it is much closer to being monotonic than any of the 2-dimensional projects. 

\begin{figure}[t!]
    \centering  
    \includegraphics[width=8cm]{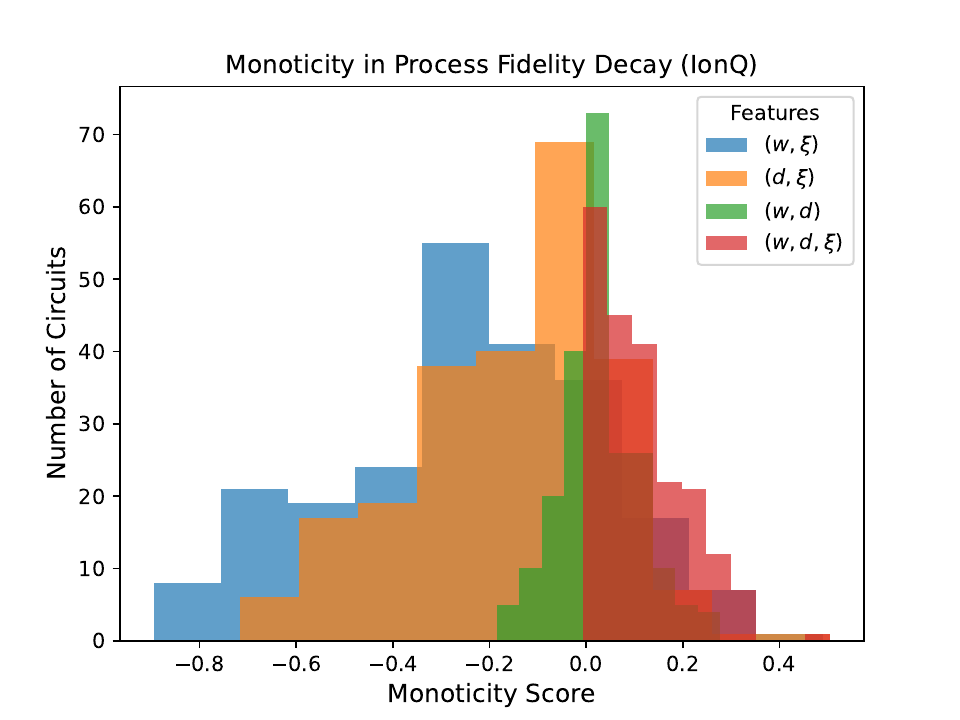}
    \caption{\textbf{Quantifying the monotonicity of the circuit fidelity decay (\texttt{Forte1})}. We quantify the monotonicity of the decay in circuit fidelities with increasing feature values using a simple metric $\delta_{\vec{v}}$ given by the minimum of $F_{\vec{v}'}- F_{\vec{v}}$ over all $\vec{v}'$ that are strictly smaller than $\vec{v}$ (i.e., all $\vec{v}'$ that are equal or smaller for every feature, and smaller for at least one feature). Negative values of $\delta_{\vec{v}}$ indicate feature values at which we observed a process fidelity that was larger for a smaller feature value, which is inconsistent with monotonicity. The data is almost monotonic when using all three features (red histogram), but substantially non-monotonic when discarding any of the three features (blue, orange, and green histograms).}
    \label{fig:ionq_monoticity}
\end{figure}

We quantify consistency with monotonicity in Fig.~\ref{fig:ionq_monoticity}, where we show a simple metric for how monotonic our dataset is, as well as how monotonic each 2-dimensional projection is. For each feature value $\vec{v} = (w,d,\xi_{\textrm{2Q}})$ with observed process fidelity $F_{\vec{v}}$, we compute the minimum $\delta_{\vec{v}}$ of $F_{\vec{v}'} - F_{\vec{v}}$ over all feature vectors $\vec{v}'=(w',d',\xi_{\textrm{2Q}}')$ that are strictly smaller than $\vec{v}$ (i.e., $w'\leq w$, $d' \leq d$, $\xi_{\textrm{2Q}} \leq \xi_{\textrm{2Q}}$ and at least one feature is strictly smaller). We do the same for each 2-dimensional project, whereby we ignore one of the three features. Fig.~\ref{fig:ionq_monoticity} shows a histogram of $\delta_{\vec{v}}$, as well as its value for the 2-dimensional projections ($\delta_{(w,d)}$, $\delta_{(w,\xi_{\textrm{2Q}})}$, and $\delta_{(d,\xi_{\textrm{2Q}})}$). Negative values of $\delta_{\vec{v}}$ correspond to feature values for which a smaller process fidelity was observed for a smaller value of the feature vector. We observe that the data is almost monotonic when using all three features (red histogram), but it is not statistically consistent with monotonicity: for some feature vectors $\delta_{\vec{v}}$ is negative and statistically inconsistent with zero (e.g., for one feature vector $\delta_{\vec{v}} = (-4 \pm 1)\%$, where the uncertainty is the standard error, so this $\delta$ is four $\sigma$ inconsistent with being non-negative, and there are 248 different events---i.e., features---so we would not expect to see a 4-sigma event by chance). However, when we discard any of the three features, we find that the data is far from consistent with monotonicity (blue, orange, and green histograms).

\begin{figure*}[t!]
    \centering  
    \includegraphics[width=18cm]{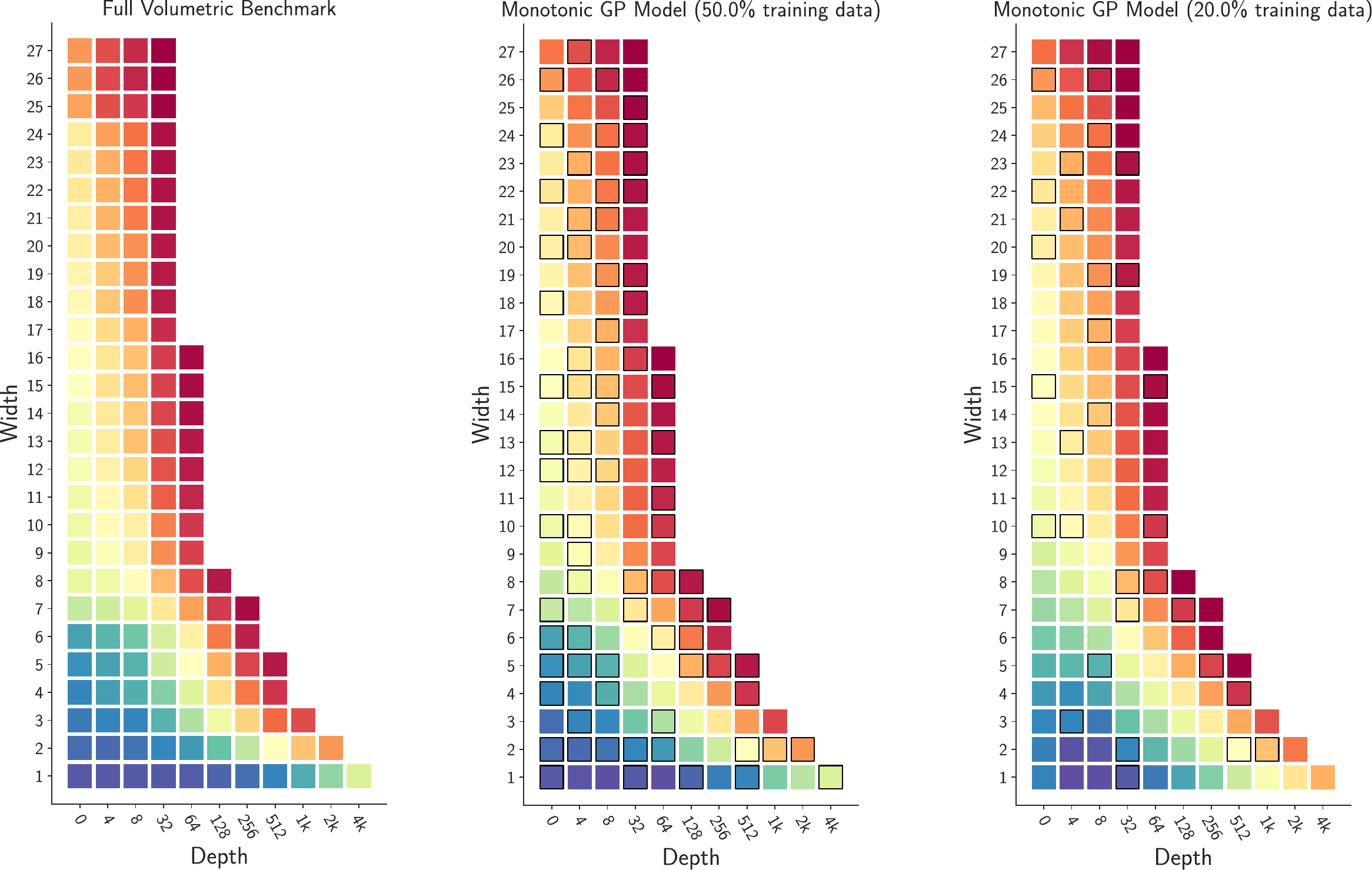}
    \caption{\textbf{Learning a volumetric model from fewer data (\texttt{ibmq\_montreal})}. The standard approach to creating a volumetric benchmarking plot (a.k.a.~capability region) requires running many circuits---$K \gg 1$ at each width and depth in a grid of circuit shapes. In this work, we show how to approximate a volumetric benchmarking plot from many fewer data, using Gaussian process (GP) regression. Here we show a full volumetric benchmarking plot for \texttt{ibmq\_montreal} (left), described in Sec.~\ref{sec:experiment-montreal} and Fig.~\ref{fig:ibm_montreal_vb}, and predictions for this plot created from only 50\% (center) and 20\% of the data (right), selected at random from the full dataset. These predictions are created by training a (monotonic) GP on that data (squares with black outlines), which creates a model that predicts circuit success probabilities at any circuit shape $(w,d)$. Here we show that model's predictions on the remaining test data, which creates an extra/interpolated volumetric benchmarking plot. The GP model that was trained on 50\% of the data is very accurate, and even the model trained on 20\% of the data is very accurate at interpolation as opposed to extrapolation (suggesting that model accuracy could be improved with more systematically chosen values for the circuit shapes in the training set).}
    \label{fig:interpolated_vb}
\end{figure*}

\section{Featuremetric capability models}\label{sec:analysis}
We now present methods for analyzing featuremetric benchmarking data, to create performance summaries and predictive capability models, and we apply these methods to the data sets described above. The techniques we present in this section are designed for general featuremetric benchmarking data, but they are also of interest for the special case of volumetric benchmarking.

 \begin{figure}[t!]
    \centering  
    \includegraphics[width=7.5cm]{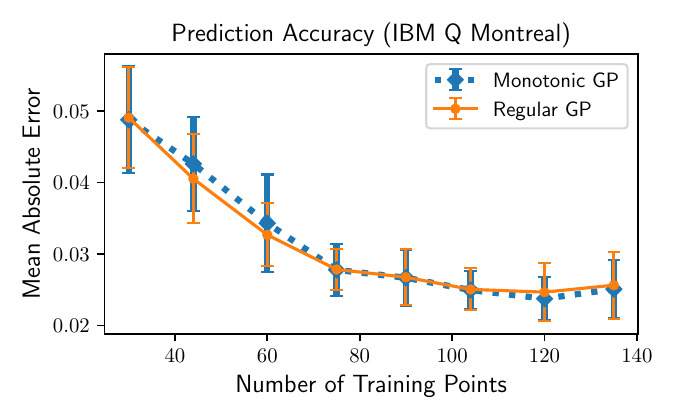}
    \includegraphics[width=7.5cm]{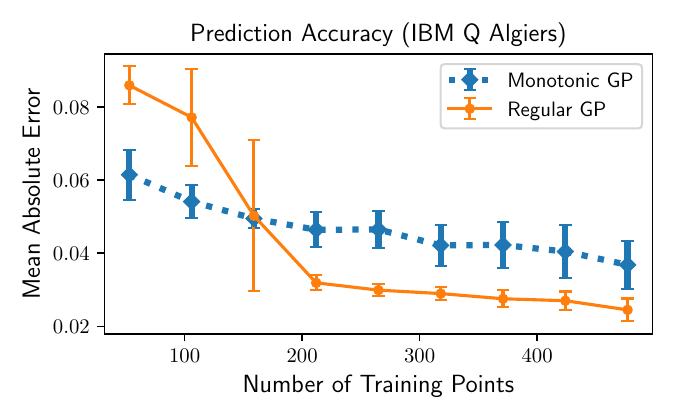}
    \includegraphics[width=7.5cm]{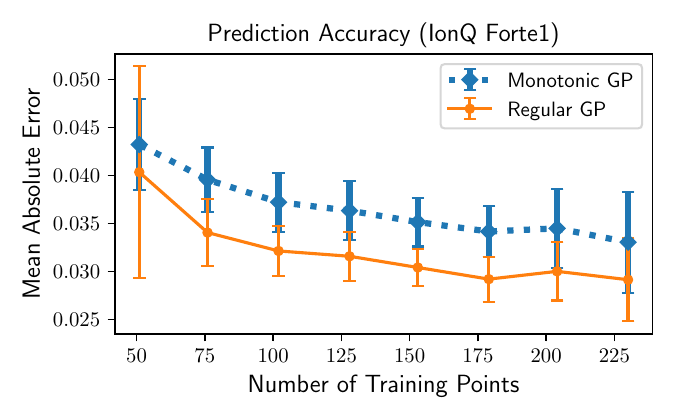}
    \caption{\textbf{Model prediction error}. The prediction error of our featuremetric models for each of our three datasets, on the test data, versus the amount of training data. We created these models using regular (orange points) and monontonic (blue diamonds) GP regression. Prediction error is quantified by the mean absolute difference $\delta$ between predicted and observed circuit performance (success probabilities for both IBM Q datasets and circuit process fidelities for the IonQ dataset), where the mean is over all feature values in the test data set. Points show the mean of $\delta$ for 20 different GP instances trained on randomly-sampled subsets of data of the specified size, and the error bars show the standard deviation of $\delta$ over those 20 instances. We find that model prediction accuracy improves as the amount of training data increases, but that, in all cases, accuracy is reasonably good when using $\sim 50\%$ or more of the data to train (a 50\% test/training split corresponds to 75, 266, and 128 training data, for \texttt{ibmq\_montreal}, \texttt{ibmq\_algiers}, and \texttt{Forte1}, respectively).}
    \label{fig:model_error}
\end{figure}

\subsection{Learning a low-dimensional representation of capabilities}
A featuremetric benchmarking experiment studies a \emph{low-dimensional} representation of a quantum computer's capability function $s(c)$. In particular, a featuremetric benchmarking experiment with feature vector $\vec{f}$ studies how $s(c)$ varies with $\vec{f}$. We now make this statement more precise. For clarity, we will first ignore the error in estimating $s(c)$ from data (due to finite repetitions of each circuit), i.e., we assume that we measure $s(c)$ exactly for each selected benchmarking circuit $c$.

In featuremetric benchmarking, we pick a circuit to run with feature values $\vec{v}$ by sampling it from a distribution $P_{\vec{v}}$ over $\mathbb{C}$, we then measure that circuit's $s(c)$, and then we (implicitly) discard all information about $c$ except $\vec{v}$. This can be formalized in the language of random variables as follows: we observe $K$ samples from the random variable
\begin{equation}
    s(\vec{v}) := s(C_{\vec{v}}),
\end{equation}
at each value for $\vec{v}$, where $C_{\vec{v}}$ is a circuit-valued random variable that is distributed according to $P_{\vec{v}}$. A featuremetric benchmarking experiment is probing the distribution of $s(\vec{v})$ as a function of $\vec{v}$. If a particular set of circuit features $\vec{f}$ are entirely predictive of $s(c)$, i.e., if given the value for a circuit $c$'s features $\vec{f}$ we can predict $s(c)$ exactly, then $s(\vec{v})$ is a delta distribution for all $\vec{v}$. However, this is unlikely to be the case in reality, for any small set of features.

The goal of the featuremetric benchmarking analysis is to learn about the parameterized distribution $s(\vec{v})$ from the featuremetric benchmarking data. This data has the form
\begin{equation}
    x = \{(\vec{v}_i, y_i)\}
\end{equation}
where
\begin{equation}
    y_i = (\hat{s}(c_{i,1}), \dots, \hat{s}(c_{i,K})),
\end{equation} 
$\hat{s}(c)$ denotes an estimate of $s(c)$ from finite data (often simple finitely-many executions of $c$), and $s(c_{i,j})$ can be interpreted as a sample from $s(\vec{v}_i)$. From this data, we aim to learn about $s(\vec{v})$. Most generally, we would like to learn an approximation to $s(\vec{v})$, i.e., we aim to learn an $\vec{v}$-parameterized distribution over $\mathbb{R}$ (or, more precisely, the image of $s$, which is $[0,1]$ for many interesting choices for $s$, such as success probability). However, learning this parameterized distribution is a challenging problem. Throughout the remainder of this paper, we will therefore focus on a simpler problem: learning an approximation to $s(\vec{v})$'s mean, which we denote by $\bar{s}(\vec{v})$, from the data $x =\{ (\vec{v}_i, \hat{\bar{s}}_i)\}$ where $\hat{\bar{s}}_i = \frac{1}{K}\sum_{j=1}^{K}\hat{s}(c_{i,j})$.

Our analysis task is a standard regression problem---learning a function $\bar{s}(\vec{v})$ from data $x$---for which many techniques exist. The conceptually simplest approach is to fit a few-parameter function (a parametric model) to the data. For example, with the two circuit features width and depth, we could fit the function $\bar{s}(w,d)=p^{wd}$ where $p$ is a fit parameter. The problem with this approach is that we do not typically have a good few-parameter model for $s\bar{s}(\vec{v})$. Furthermore, one of the most appealing aspects of benchmarks like featuremetric benchmarking is that (unlike, e.g., tomography) they do not need to assume a particular model for quantum computer's errors. Returning to our example, $\bar{s}(w,d)=p^{wd}$ is a good model if we know that all error processes are uniform, gate-independent, $n$-qubit depolarization, but this is unlikely to be true for any contemporary quantum computer. Furthermore, if we are \emph{a priori} confident that about the correctness of this model, or some other simple model, there is no need to run a featuremetric benchmarking experiment. We can instead just learn $p$'s value with some simpler experiments (e.g., randomized benchmarking).

The alternative approach, that we take in this work, is to use non-parametric regression methods, which do not assume a specific, few-parameter model. There are many such possibilities, such as neural networks, and in this work we focus on GP (Gaussian process) regression \cite{rasmussen2006gaussian}. GP regression is a non-parametric Bayesian machine learning approach, used for predicting the values of a function based on observed data, where the function is assumed to be a realization of a Gaussian process. It provides a rigorous framework for incorporating uncertainty in predictions by defining a prior distribution over functions and updating this prior with observed data to obtain a posterior distribution. Because we use non-parametric methods, to assess the quality of our models we split our data into test and training data, fitting the models to the training data and testing their prediction accuracy on the test data. We provide a mathematical overview of GP regression in Appendix~\ref{app:regular-gpr}.

We do not typically have a good few-parameter functional form for $\bar{s}(\vec{v})$ but for many interesting choices of $\vec{v}$ we have good reasons to expect that $\bar{s}(\vec{v})$ has certain structures. In particular, for many features $v_i$ we expect $\bar{s}(\vec{v})$ to monotonically increase or decrease as $v_i$ increases. For example, for typical circuit families, we expect  $\bar{s}$ to decrease with increasing circuit depth, width, and, in systems in which two-qubit gate error rates are significantly larger than single-qubit gate error rates, two-qubit gate density. We can incorporate this assumption into our model using a \emph{monotonic} GP \cite{riihimaki2010gaussian, vanhatalo2012bayesian, vanhatalo2013gpstuff}. Monotonic GP regression learns a function that is approximately monotonic. Throughout this section we apply both regular GP regression and monotonic GP regression to featuremetric benchmarking data. We provide a mathematical overview of monotonic GP regression in Appendix~\ref{app:monotonic-gpr}.

 \begin{figure}[t!]
    \centering  
    \includegraphics[width=8.5cm]{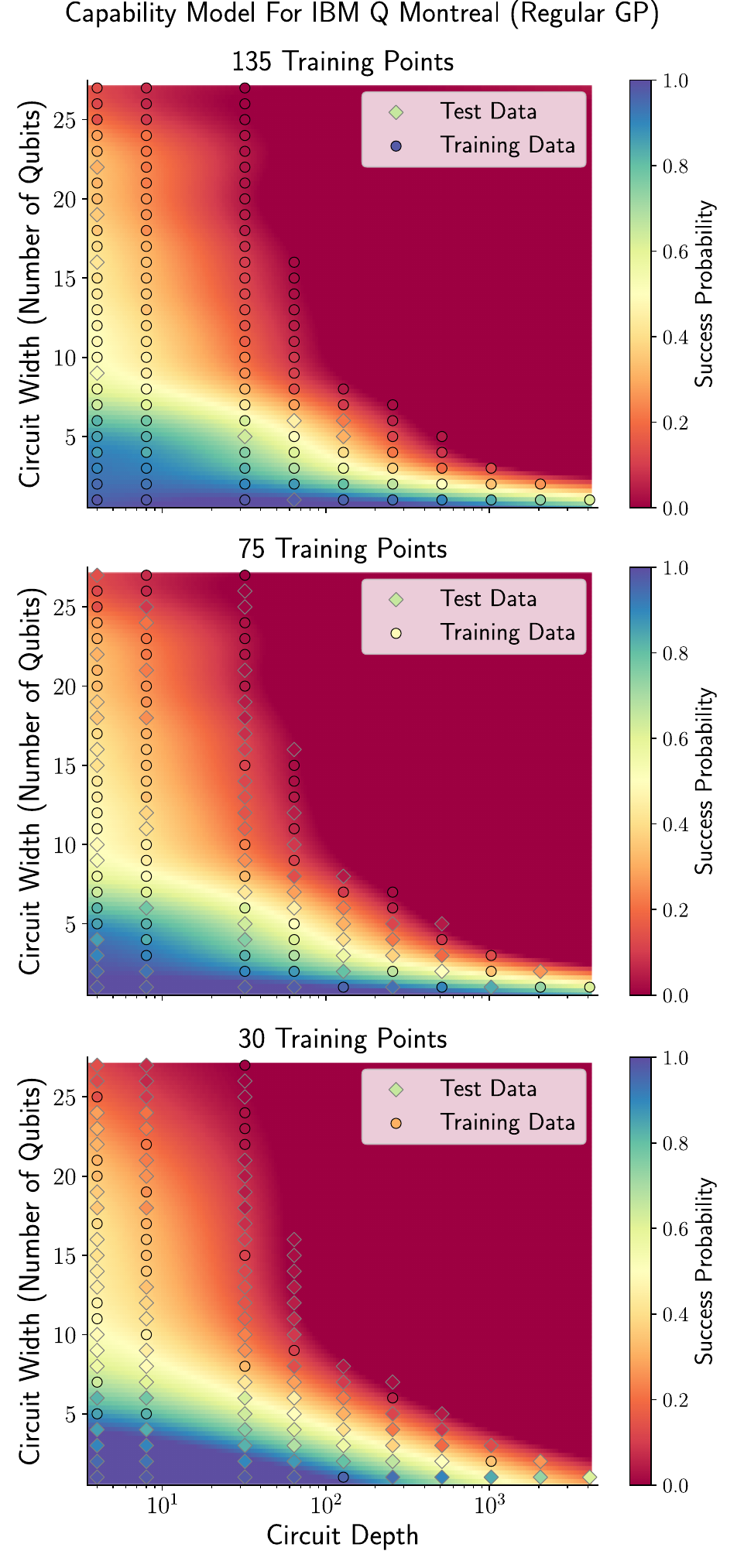}
    \caption{\textbf{Continuous volumetric benchmarking plots.} Capability models enable the prediction of a circuit's performance with arbitrary values for its features, i.e., the capability model $\mathcal{M}$ can take any value for the feature vector and makes a prediction $\mathcal{M}(\vec{v})$ for the performance of a circuit with that feature vector. In the case of volumetric benchmarking data, the capability model's predictions can be represented as a ``continuous volumetric benchmarking plot''. Here we show these predictions for our \texttt{ibmq\_montreal} dataset, with three different models trained on varying amounts of the data. We also show the training (circles) and test (diamonds) data.}
    \label{fig:montreal_gpr_heatmap}
\end{figure}

 \begin{figure*}[t!]
    \centering  
    \includegraphics[width=5.5cm]{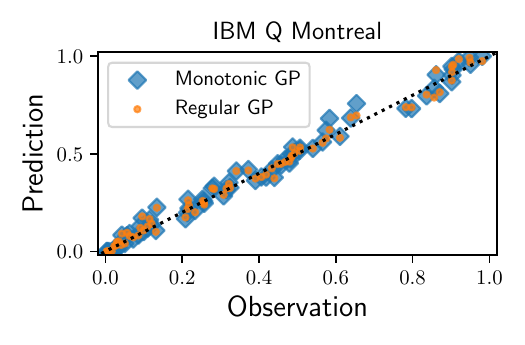}
    \includegraphics[width=5.5cm]{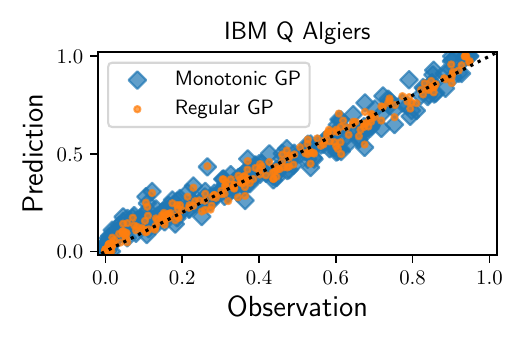}
    \includegraphics[width=5.5cm]{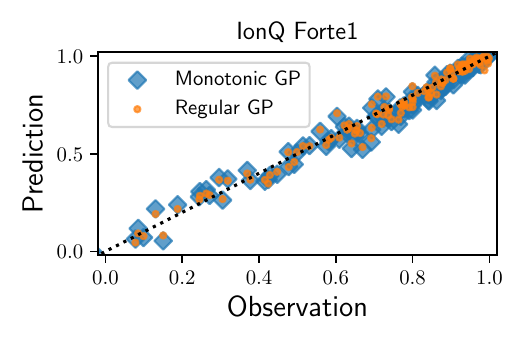}
    \caption{\textbf{Model predictions on test data}. The prediction error of our featuremetric models for each of our three datasets, on the test data, when using 50\% of the data to train each GP (this is for a single instance with one particular random division of the data into test and training sets). In all three cases, we find reasonable prediction accuracy.}
    \label{fig:model_predictions_50p}
\end{figure*}

\subsection{Learning capability models with GP regression}
We used regular and monotonic GP regression to learn capability models from each of our featuremetric benchmarking datasets. In each case, our model $\mathcal{M}$ predicts $\bar{s}(\vec{v})$, i.e., it is a function from a feature vector $\vec{v}$ to a prediction for the mean value of the capability function at that feature value. To train our model $\mathcal{M}$ we randomly divide our data into test and training subsets, with the model fit to the training data only, which is standard practice and allows us to assess model accuracy on independent data. To explore how much data we need to construct accurate capability models, we varied the amount of training data.

Capability models for two-dimensional feature vectors are easiest to visualize and understand, and so we first focus on our \texttt{ibmq\_montreal} dataset. Such capability models can be interpreted as an interpolation and extrapolation of volumetric benchmarking data. In Fig.~\ref{fig:interpolated_vb} we show the volumetric benchmarking plot for this data (left panel) as well as two volumetric plots (center and right panels) that are predicted from a GP that is trained on a subset of the data. We train a GP on two randomly-selected subsets, containing 50\% and 20\% of the data points, and then predicted the mean success probabilities for the other $(w,d)$ values in the volumetric benchmarking plot. We find that both predicted volumetric benchmarking plots are in good agreement with the measured plot, with increasing predicting accuracy as the amount of training data increases. This shows how GP regression can be used to create a volumetric benchmarking plot from many fewer data.

To quantify the quality of the prediction accuracy of the learned capability model, we compute the mean absolute difference between the predicted and observed mean success probabilities $\delta_{\textrm{abs}}$ for the test data (for this dataset, this means comparing the predicted values on the test data on the volumetric benchmarking grid). In the upper panel of Fig.~\ref{fig:model_error} we show $\delta_{\textrm{abs}}$ versus the amount of training data, for both a monotonic and a regular GP. Points show the mean of $\delta_{\textrm{abs}}$ averaged over 20 different randomly-selected divisions into training and test datasets, and error bars are the standard deviation in $\delta_{\textrm{abs}}$ over those 20 instances. We find that the model's prediction error decreases with increasing training dataset size, for both the monotonic and regular GP, converging to $\delta_{\textrm{abs}} \approx 3\%$ once around $75$ training data are used.

One of the appealing aspects of a learned capability model is that it can make predictions at \emph{any} feature vector value, not just those that were used in the experiment. In particular, it can be used to produce a prediction for the mean success probability of circuits of any width and depth. For two-dimensional feature vectors, the model's predictions can therefore be summarized in a heat map---a ``continuous volumetric plot''---as shown in Fig.~\ref{fig:montreal_gpr_heatmap} for three example models (regular GPs trained on data from 30, 75, and 135 feature values).

Our capability models for our two three-dimensional data sets are more complex to visualize, but can be just as easily used to predict the performance of any circuit given the values of that circuit's feature vector---in this case, a three-dimensional vector, consisting of the circuit's width, depth, and two-qubit gate density. In the middle and lower panels of Fig.~\ref{fig:model_error} we show $\delta_{\textrm{abs}}$ versus the amount of training data for these two datasets, for both a monotonic and regular GP trained on varying amounts of the datasets. This figure summarizes each model's accuracy with a single figure of merit, which averages over all the test data. So, in Fig.~\ref{fig:model_predictions_50p}, we show the prediction accuracy for each test feature value, for all three datasets, when using 50\% of the data to train each GP (this is for a single instance with one particular random division of the data into test and training sets). Figures~\ref{fig:model_error} and~\ref{fig:model_predictions_50p} show that the GP models exhibit good prediction accuracy on the test data, suggesting that GP models will enable reasonably good predictions of circuit performance.

\section{Conclusions}\label{sec:discussion}
We have presented and demonstrated \emph{featuremetric benchmarking}---a framework for quantifying how a quantum computer's performance on quantum circuits varies as a function of \emph{features} of those circuits, such as circuit depth, width, two-qubit gate density, or problem input size. This framework generalizes volumetric benchmarking, and it enables richer and more faithful models of quantum computers' performance. One key requirement for informative featuremetric benchmarking is a well-chosen set of circuit features to vary, and in this work we only consider three such features: width, depth, and two-qubit gate density. The development of a richer and well-motivated set of features would likely enhance featuremetric benchmarking, by enabling the learning of capability models that are more accurate predictors of circuit performance. One intriguing possibility is to use machine learning methods to learn pertinent features of circuits \cite{Hothem2024-rc, Hothem2024-fc}.

We demonstrated featuremetric benchmarking with example benchmarks run on IBM Q and IonQ systems of up to 27 qubits, and we showed how to create capability models from the data using GP regression. Our GP models enable the prediction of any circuit's performance from featuremetric benchmarking data, although it unlikely that our current feature sets are rich enough for our GP models to accurate predict the performance of circuits that are not drawn from the same distribution as the training data. One simple but interesting use for our GP models is the creation of volumetric benchmarking plots from many fewer data, as, when applied to volumetric benchmarking data (the circuit features of width and depth) they enable the predictions of circuit performance at any other width and depth---and we demonstrated how a volumetric benchmarking plot could be predicted with reasonable accuracy using just 20\% of its data.

Our GP models demonstrate how a moderately accurate capability model for a quantum computer can be learned with relatively few data. However, our GP regression approach can likely be improved in a variety of ways. Model training is not currently automated and stable, and our approach of randomly selecting training data is likely far from optimal. In particular, our training data selection method can result in scenarios in which the training data is very unevenly distributed across the feature space, and where the prediction of the test data is an extrapolation rather than interpolation problem (reliable extrapolation is typically a much more difficult problem than interpolation). One particularly appealing aspect of using ML techniques like GP regression to learn capability models, rather than simply plotting data as in volumetric benchmarking, is that we can reduce the amount of data needed---and therefore the experimental cost---to benchmark a quantum computer. This improved efficiency could likely be further enhanced using \emph{online} learning methods, whereby a GP is trained on a small amount of initial data and then it is used to select what feature values to explore next. GPs are particularly well-suited to this approach, as they are a Bayesian method that produces a posterior distribution and they incorporate uncertainty on their predictions.

\section*{Acknowledgments}
This material is based upon work supported by the U.S. Department of
Energy, Office of Science (DE-FOA-0002253), National Quantum Information Science Research Centers, Quantum Systems Accelerator. Sandia National Laboratories is a multi-program laboratory managed and operated by National Technology and Engineering Solutions of Sandia, LLC., a wholly owned subsidiary of Honeywell International, Inc., for the U.S. Department of Energy's National Nuclear Security Administration under contract DE-NA-0003525. All statements of fact, opinion or conclusions contained herein are those of the authors and should not be construed as representing the official views or policies of the U.S. Department of Energy or the U.S. Government.

\bibliography{bibliography}

\appendix
\section{SPAM-error-robust DFE}\label{app:dfe}
In this appendix we explaining \emph{spam-error-robust directly fidelity estimation} (SR-DFE) \cite{Seth2025-zz}, which is the technique used in our IonQ experiments. SR-DFE is an adaptation of DFE, and so we first review DFE. For an $n$-qubit Clifford circuit $c$, DFE estimates $F(c)$ by running $K$ circuits each consisting of
\begin{enumerate}
  \item a layer of randomly-sampled single-qubit gates applied to the standard initial state (i.e., $|0\rangle\otimes \cdots \otimes |0\rangle$),
  \item $c$, and
  \item another layer of single-qubit gates followed by a computational basis measurement. 
\end{enumerate}
 The first layer of single-qubit gates is sampled as follows: a \emph{uniformly random} $n$-qubit Pauli operator $P_1$ is sampled, and then this layer of gates is chosen so that it creates a state $|\psi_1\rangle$ from $|0\cdots0\rangle$ that is stabilized by $P_1$, i.e., \begin{equation}
     P_1|\psi_1\rangle=|\psi_1\rangle.
 \end{equation}
 The circuit $c$ transforms $|\psi_1\rangle$ into another stabilizer state $|\psi_2\rangle = U|\psi_1 \rangle$ that is stabilized by $P_2=UP_1U^{\dagger}$ where $U$ is the unitary implemented by $c$. The final layer of single-qubit gates is then chosen to transform $P_2$ into a $Z$-type Pauli operator $P_3$---i.e., $P_3$ is an $n$-qubit Pauli operator that is a tensor product of only $I$ and $Z$. The final state produced by an error-free execution of the circuit, denoted $|\psi_3 \rangle$, is stabilized by $P_3$, i.e., 
 \begin{equation}
 P_3|\psi_3\rangle=|\psi_3\rangle,
 \end{equation}
 and $P_3$ can be measured using a computational basis measurement.
 
 The DFE analysis consists of estimating $\langle P_3 \rangle$ from $N \geq 1$ repetitions of the circuit: each repetition of the circuit returns a bit string $b$ that is associated with either a $+1$ or $-1$ eigenstate of $P_3$, i.e., either $P_3|b\rangle=|b\rangle$ or $P_3|b\rangle=-|b\rangle$. Denoting the set of all $x$ such that $P_3|x\rangle=|x\rangle$ by $\mathbb{Z}_{+}$, we estimate $\langle P_3 \rangle$ as
\begin{equation}
    \widehat{\langle P_3 \rangle} =  \frac{N(x \in \mathbb{Z}_+) - N(x \notin \mathbb{Z}_+)}{N},
\end{equation}
where $N(x \in \mathbb{Z}_+)$ and $N(x \notin \mathbb{Z}_+)$ are the number of times the circuit's output $x$ is and is not in $\mathbb{Z}_{+}$, respectively. DFE then estimates $F(c)$, by averaging $\widehat{\langle P_3 \rangle}$ over the $K$ circuits (corresponding to $K$ different randomly sampled $P_1$).

DFE reliably estimates $F(c)$ under the assumption that the SPAM and layers of single-qubit gates (steps 1 and 3, above) are free of errors \cite{Flammia2011-qj,Moussa2012-rq, Hashim2024-om}. In practice, this assumption is always violated (SPAM errors are significant in many systems), which corrupts DFE's estimate of $F(c)$. SR-DFE is a simple adaptation to DFE that makes it (approximately) robust against these SPAM errors. SR-DFE estimates $F(c)$ by:
\begin{enumerate}
\item Using DFE to obtain an estimate of $F(c)$, which we denote by $\hat{F}_{\textrm{DFE}}(c)$. In general this estimate is corrupted by SPAM errors.
\item Using DFE on a trivial circuit $c_{\textrm{null}}$ (a depth-0 circuit, containing no layers), to obtain an estimate of the fidelity of $F(c_{\textrm{null}})$, which we denote by $\hat{F}_{\textrm{DFE}}(c_{\textrm{null}})$.
\end{enumerate}
The latter DFE procedure estimates the fidelity of the effective SPAM operations (meaning all operations in steps 1 and 3), which can then be removed from $\hat{F}_{\textrm{DFE}}(c)$. In particular, the SR-DFE estimate of $F(c)$ is given by
\begin{equation}
 \hat{F}(c) =  \Gamma_{\gamma \to F} \left(\frac{\Gamma_{F \to \gamma}( \hat{F}_{\textrm{DFE}}(c) , n)}{\Gamma_{F \to \gamma}(\hat{F}_{\textrm{DFE}}(c_{\textrm{null}}), n)}, n \right)\approx  \frac{\hat{F}_{\textrm{DFE}}(c)}{\hat{F}_{\textrm{DFE}}(c_{\textrm{null}})}
\end{equation}
where $\Gamma_{F \to \gamma}(F,n)$ is the function that computes the \emph{process polarization} for an $n$-qubit superoperator from its process fidelity $F$ and $n$, and $\Gamma_{\gamma \to F}$ is its inverse. Specifically,
\begin{equation}
\Gamma_{F \to \gamma}(F, n) = \frac{4^n F - 1}{4^n - 1}.
\end{equation}

The reason that we do not estimate $ \hat{F}(c)$ by simply dividing $\hat{F}_{\textrm{DFE}}(c)$ by $\hat{F}_{\textrm{DFE}}(c_{\textrm{null}})$ is that, for two Pauli stochastic superoperators $A$ and $B$, the process polarization of $AB$ is approximately equal to the product of the two superoperators process polarizations, but for very small $n$ this is not true of process fidelities. This is identical to the reasoning used in the analysis of MCFE, and it is discussed in detail in Ref.~\cite{Proctor2022-zs}.

In our experiments on IonQ we use SR-DFE to estimate the process fidelity of many different circuits $\{c\}$. These circuits vary in their width, and so in the qubits on which they are executed, but there are many circuits in $\{ c\}$ that act on the same qubits. For any two circuits acting on the same qubits, the procedure for generating the SPAM reference circuits (step 2 in the SR-DFE procedure, above) is identical. In our experiments, we run one SPAM reference circuits for each circuit $c$, directly after the DFE circuits for $c$ (step 1 in the SR-DFE procedure), and use the $K$ SPAM reference circuits run for the $K$ circuits at a given feature vector value $\vec{v}$ to correct the estimates of the fidelities for those $K$ circuits. This procedure adds robustness to drift in the size of the effective SPAM error. We use only one DFE circuit (together with the SPAM reference circuits) to estimate the process fidelity of each circuit $c$. This gives a low-precision estimate of that individual circuit's process fidelity, but in our analysis herein we only study the mean process fidelity of all $K$ circuits chosen for each feature vector.

\section{Gaussian process regression}\label{app:regular-gpr}

Gaussian process regression is a non-parametric Bayesian machine learning approach used for predicting the values of a function based on observed data, where the function is assumed to be a realization of a Gaussian process. It provides a mathematically rigorous framework for incorporating uncertainty in forward predictions by defining a prior distribution over functions and updating this prior with observed data to obtain a posterior distribution. 

Let us assume that we can model the true process, $\mathbf{y}$, with a zero-mean GP,
\begin{equation}
\mathbf{f} ( \mathbf{x}) \sim \mathcal{N}(\mathbf{0}, \mathbf{K}_{\mathbf{f},\mathbf{f}}),
\end{equation}
where the entries in the covariance matrix $\mathbf{K}_{\mathbf{f}, \mathbf{f}}$ are defined by a covariance function.
Although there are many covariance functions to choose from, in this paper, we focus on the squared exponential covariance function
\begin{equation}
\Cov{ f^{(i)} , f^{(j)} } = \mathbf{K}(\mathbf{x}^{(i)}, \mathbf{x}^{(j)}) = \eta^2 \exp\left[ - \frac{1}{2} \sum_{d=1}^D \rho_d^{-2} (x^{(i)}_d - x^{(j)}_d)^2 \right],
\label{eq:kernel}
\end{equation}
where $\eta$ and $\mathbf{\rho} = \{\rho_1, \dots, \rho_D \}$ are hyper-parameters, representing the signal variance and length-scale (also known as correlation length) parameters, respectively.
The observations $\mathbf{y}$ are then given by
\begin{equation}
\mathbf{y} | \mathbf{f} \sim \mathcal{N}(\mathbf{f}, \eta^2 \mathbf{I}),
\end{equation}
where $\eta^2$ is the intrinsic noise variance. 

Let the training dataset be denoted as $ \mathbf{X} = \{\mathbf{x}_i\}_{i=1}^n $, representing $n$ data points. At an arbitrary test location \( x^* \), the posterior predictive distribution - also known as the testing distribution - is Gaussian, where the posterior mean and the posterior variance given by
\begin{equation}
\E{f^* | x^*, \mathbf{y}, \mathbf{X}, \mathbf{\theta}} = \mathbf{K}_{\mathbf{*},\mathbf{f}} [\mathbf{K}_{\mathbf{f},\mathbf{f}} + \sigma^2 \mathbf{I}]^{-1} \mathbf{y}
\end{equation}
and
\begin{equation}
\V{f^* | x^*, \mathbf{y}, \mathbf{X}, \mathbf{\theta}} = \mathbf{K}_{\mathbf{*},\mathbf{*}} - \mathbf{K}_{*,\mathbf{f}} [\mathbf{K}_{\mathbf{f},\mathbf{f}} + \sigma^2 \mathbf{I}]^{-1} \mathbf{K}_{\mathbf{f},\mathbf{*}},
\end{equation}
respectively. 
The covariance matrix $\mathbf{K}_{\mathbf{f},\mathbf{f}} \in \mathbb{R}^{n \times n}$ is a symmetric positive definite matrix 
Here, the vector of hyper-parameter is $\mathbf{\theta} = \{ \eta, \rho_1, \dots, \rho_D\}$, with one length-scale parameter $\rho_d$ for each dimension $d$, $1 \leq d \leq D$. 
GPR is trained by maximizing the log-marginal likelihood with respect to the vector of hyper-parameters as
\begin{equation}
\begin{array}{lll}
\log p(\mathbf{y} | \mathbf{X}, \mathbf{\theta}) &=& - \frac{1}{2}\mathbf{y}^\top [\mathbf{K}_{\mathbf{f},\mathbf{f}} + \sigma^2 \mathbf{I}]^{-1} \mathbf{y} \\
&& - \frac{1}{2} \log{|\mathbf{K}_{\mathbf{f}, \mathbf{f}} + \sigma^2 \mathbf{I}|} - \frac{N}{2} \log{(2\pi)}
\end{array}
\label{eq:mle}
\end{equation}
In \ref{eq:mle}, the first term, known as the ``data fit'' term, measures how well the model represents the data in the Mahalanobis distance. The second term, referred to as the ``complexity'' term, captures model complexity, favoring smoother covariance matrices with smaller determinants~\cite{rasmussen2006gaussian}. The final term reflects the tendency for data likelihood to decrease as the dataset size increases~\cite{shahriari2016taking}. 
The complexity of Equation~\ref{eq:mle} is $\mathcal{O}(n^3)$ in computation, due to the determinant calculation and covariance matrix inversion, and $\mathcal{O}(n^2)$ in memory. As a result, GPR is generally limited to small datasets, typically with $n$ fewer than $10^4$ data points.

Thanks to the linear property of the expectation operator, we have that
\begin{equation}
\E{ \frac{\partial f^{(i)}}{\partial x_d^{(i)}} } =  \frac{\partial \E{f^{(i)}}}{\partial x_d^{(i)}},
\end{equation}
\begin{equation}
\Cov{ \frac{\partial f^{(i)}}{\partial x_d^{(i)}} , f^{(j)} } =  \frac{\partial }{\partial x_d^{(i)}} \Cov{f^{(i)}, f^{(j)}},
\label{eq:covFirstDerivative}
\end{equation}
and
\begin{equation}
\Cov{ \frac{\partial f^{(i)}}{\partial x_d^{(i)}} , \frac{\partial f^{(j)}}{\partial x_g^{(j)}} } =  \frac{\partial^2 }{\partial x_d^{(i)} \partial x_g^{(j)} } \Cov{f^{(i)}, f^{(j)}}.
\label{eq:covSecondDerivative}
\end{equation}
For the squared exponential kernel described in Equation~\ref{eq:kernel}, Equations~\ref{eq:covFirstDerivative} and~\ref{eq:covSecondDerivative} become
\begin{equation}
\begin{array}{lll}
\Cov{ \frac{\partial f^{(i)}}{\partial x_g^{(i)}} , f^{(j)} } &=& - \eta^2 \exp\left( -\frac{1}{2} \sum_{d=1}^D \rho_d^{-2} (x_d^{(i)} - x_d^{(j)})^2 \right) \\
&& \rho_g^{-2} \left( (x_g^{(i)} - x_g^{(j)}) \right) ,
\end{array}
\end{equation}
and
\begin{equation}
\begin{array}{lll}
\Cov{ \frac{\partial f^{(i)}}{\partial x_d^{(i)}} , \frac{\partial f^{(j)}}{\partial x_h^{(j)}} } &=& \eta^2 \exp\left( -\frac{1}{2} \sum_{d=1}^D \rho_d^{-2} (x_d^{(i)} - x_d^{(j)})^2 \right) \\
&& \rho_g^{-2} \left( \delta_{gh} - \rho_h^{-2} (x_g^{(i)} - x_g^{(j)}) (x_h^{(i)} - x_h^{(j)}) \right) ,
\end{array}
\end{equation}
respectively, where $\delta_{gh} = 1$ if $g=h$ and 0 otherwise.

For an arbitrary testing point $x^*$, the derivatives with respect to the dimension of the posterior mean and posterior variance are, respectively,
\begin{equation}
\E{ \frac{\partial f^*}{\partial x^*_d} } = \frac{\partial \mathbf{K}_{*, \mathbf{f}}}{\partial x^*_d} [\mathbf{K}_{\mathbf{f}, \mathbf{f}} + \sigma^2 \mathbf{I}]^{-1} \mathbf{y},
\end{equation}
and
\begin{equation}
\begin{array}{lll}
\V{ \frac{\partial f^*}{\partial x^*_d} } =  \frac{\partial^2 \mathbf{K}_{*, *}}{\partial x^*_d \partial x^*_d} - \frac{\partial \mathbf{K}_{*, \mathbf{f}}}{\partial x^*_d} [\mathbf{K}_{\mathbf{f}, \mathbf{f}} + \sigma^2 \mathbf{I}]^{-1} \frac{\partial \mathbf{K}_{\mathbf{f}, *}}{\partial x^*_d}.
\end{array}
\end{equation}

\section{Monotonic Gaussian process regression}\label{app:monotonic-gpr}

In this study, we employ the monotonic Gaussian process (GP) framework originally proposed by Riihim\"{a}ki and Vehtari~\cite{riihimaki2010gaussian}, whose implementation is available through the GPstuff toolbox~\cite{vanhatalo2012bayesian,vanhatalo2013gpstuff} and our recent works~\cite{tran2022monotonic,tran2022integrated}. To enforce monotonicity in GP regression, their approach augments the standard covariance matrix with a block structure, analogous to those used in multi-fidelity and gradient-enhanced GPs~\cite{tran2020smfbo2cogp,yang2020bifidelity,xiao2018extended}. Expectation propagation (EP)~\cite{minka2001expectation} is employed to approximate the posterior distribution, as a computationally efficient alternative to Laplace's approximation. For simplicity, we consider a zero-mean GP prior and briefly outline the formulation, directing readers to the original work~\cite{riihimaki2010gaussian} for further detail.

The method introduces monotonicity through an augmented covariance matrix \(\mathbf{K}\), where binary constraints on the derivative sign are imposed at selected input locations. The underlying idea is to pose the derivative constraint as a probabilistic classification problem, leveraging the cumulative distribution function \(\Phi(z) = \int_{-\infty}^t \mathcal{N}(t | 0,1) dt\), which arises in logistic and probit regression models.

Monotonicity is enforced at \(M\) inducing points \(\mathbf{X}_m \in \mathbb{R}^{M \times D}\). At each point \(\mathbf{x}^{(i)} \in \mathbf{X}_m\), the partial derivative of the function \(\mathbf{f}\) with respect to input \(d_i\) is required to be non-negative. This constraint is incorporated using the probit likelihood:
\begin{equation}
p\left( m_{d_i}^{(i)} \Bigg| \frac{\partial f^{(i)}}{\partial x_{d_i}^{(i)}} \right) = \Phi\left( \frac{1}{\nu} \frac{\partial f^{(i)}}{\partial x_{d_i}^{(i)}} \right),
\label{eq:probitMonotonicity}
\end{equation}
where
\begin{equation}
\Phi(z) = \frac{1}{2} \left[ 1 + \erf\left( \frac{z}{\sqrt{2}} \right) \right] = \int_{-\infty}^{z} \mathcal{N}(t | 0,1) dt
\end{equation}
is the standard normal cumulative distribution function. This likelihood is analogous to binary classification with GPs~\cite{rasmussen2006gaussian,kuss2005assessing}, with the parameter \(\nu\) controlling classification sharpness. A value of \(\nu = 10^{-6}\) is typically used.

Assuming that the function \(\mathbf{f}\) is monotonic at \(\mathbf{X}_m\), the joint prior over function values and their derivatives is:
\begin{equation}
p(\mathbf{f}, \mathbf{f}' | \mathbf{X}, \mathbf{X}_m) = \mathcal{N} \left(
\mathbf{f}_{\text{joint}} |  \mathbf{0}, \mathbf{K}_{\text{joint}}
\right),
\end{equation}
where
\begin{equation}
\mathbf{f}_{\text{joint}} = \begin{bmatrix} \mathbf{f} \\ \mathbf{f}' \end{bmatrix},
\quad
\mathbf{K}_{\text{joint}} = \begin{bmatrix} \mathbf{K}_{\mathbf{f}, \mathbf{f}} & \mathbf{K}_{\mathbf{f}, \mathbf{f}'} \\ \mathbf{K}_{\mathbf{f}', \mathbf{f}} & \mathbf{K}_{\mathbf{f}', \mathbf{f}'} \end{bmatrix}.
\end{equation}
The corresponding joint posterior is then given by Bayes' rule:
\begin{equation}
p(\mathbf{f}, \mathbf{f}' | \mathbf{y}, \mathbf{m}) = \frac{1}{Z} p(\mathbf{f}, \mathbf{f}' | \mathbf{X}, \mathbf{X}_m) p(\mathbf{y} | \mathbf{f}) p(\mathbf{m} | \mathbf{f}'),
\label{eq:truePosterior}
\end{equation}
where
\begin{equation}
p(\mathbf{m} | \mathbf{f}') = \prod_{i=1}^M \Phi\left( \frac{1}{\nu} \frac{\partial f^{(i)}}{\partial x_{d_i}^{(i)}} \right).
\end{equation}

As the posterior is not analytically tractable, we use expectation propagation to approximate it:
\begin{equation}
\begin{array}{lll}
p(\mathbf{f}, \mathbf{f}' | \mathbf{y}, \mathbf{m}) &\approx& q(\mathbf{f}, \mathbf{f}' | \mathbf{y}, \mathbf{m}) \\
&=& \frac{1}{Z_{\text{EP}}} p(\mathbf{f}, \mathbf{f}' | \mathbf{X}, \mathbf{X}_m) p(\mathbf{y} | \mathbf{f}) \prod_{i=1}^M t_i(\tilde{Z}_i, \tilde{\mu}_i, \tilde{\sigma}^2_i),
\end{array}
\label{eq:approxPosterior}
\end{equation}
where \(t_i(\tilde{Z}_i, \tilde{\mu}_i, \tilde{\sigma}^2_i) = \tilde{Z}_i \mathcal{N}(f'_i | \tilde{\mu}_i, \tilde{\sigma}^2_i)\) are local approximations obtained from EP.

This yields a closed-form Gaussian approximation:
\begin{equation}
q(\mathbf{f}, \mathbf{f}' | \mathbf{y}, \mathbf{m}) = \mathcal{N}(\mathbf{f}_\text{joint} | \mathbf{\mu}, \Sigma),
\end{equation}
where
\begin{equation}
\mathbf{\mu} = \Sigma \tilde{\Sigma}_{\text{joint}}^{-1} \tilde{\mathbf{\mu}}_{\text{joint}},
\quad
\Sigma = [\mathbf{K}_{\text{joint}}^{-1} + \tilde{\Sigma}_{\text{joint}}^{-1}]^{-1},
\end{equation}
and
\begin{equation}
\tilde{\mathbf{\mu}}_\text{joint} = \begin{bmatrix} \mathbf{y} \\ \tilde{\mathbf{\mu}} \end{bmatrix},
\quad
\tilde{\Sigma}_\text{joint} = \begin{bmatrix} \sigma^2 \mathbf{I} & 0 \\ 0 & \tilde{\Sigma} \end{bmatrix},
\end{equation}
with \(\tilde{\mu}\) being the vector of site means and \(\tilde{\Sigma} = \text{Diag}[\tilde{\sigma}_i^2]_{i=1}^M\).

The EP approximation to the log marginal likelihood is given by:
\begin{equation}
\begin{array}{lll}
\log Z_\text{EP} &=& -\frac{1}{2} \log |\mathbf{K}_\text{joint} + \tilde{\Sigma}_\text{joint}| \\
&& - \frac{1}{2} \tilde{\mathbf{\mu}}_\text{joint}^\top [\mathbf{K}_\text{joint} + \tilde{\Sigma}_\text{joint}]^{-1} \tilde{\mathbf{\mu}}_\text{joint} \\
&& + \sum_{i=1}^M \frac{(\mu_{-i} - \tilde{\mu}_i)^2}{2 (\sigma^2_{-i} + \tilde{\sigma}_i^2)} \\
&& + \sum_{i=1}^M \log \Phi\left( \frac{\mu_{-i}}{\nu \sqrt{1 + \sigma^2_{-i}/\nu^2}} \right) \\
&& + \frac{1}{2} \sum_{i=1}^M \log(\sigma^2_{-i} + \tilde{\sigma}^2_i),
\end{array}
\end{equation}
where \(\mu_{-i}\) and \(\sigma^2_{-i}\) refer to the cavity distributions in the EP update.

The inclusion of \(M\) monotonicity-inducing locations increases the computational cost from \(\mathcal{O}(N^3)\) to \(\mathcal{O}((N+M)^3)\). The predictive posterior mean and variance at a test point \(x^*\) are:
\begin{equation}
\E{f^* | x^*, \mathbf{y}, \mathbf{X}, \mathbf{m}, \mathbf{X}_m} = \mathbf{K}_{*,\text{joint}} [\mathbf{K}_{\text{joint}} + \tilde{\Sigma}_\text{joint}]^{-1} \tilde{\mu}_{\text{joint}}
\label{eq:posteriorMeanMonotonic}
\end{equation}
and
\begin{equation}
\V{f^* | x^*, \mathbf{y}, \mathbf{X}, \mathbf{m}, \mathbf{X}_m} = \mathbf{K}_{*,*} - \mathbf{K}_{*,\text{joint}} [\mathbf{K}_{\text{joint}} + \tilde{\Sigma}_\text{joint}]^{-1} \mathbf{K}_{*,\text{joint}}.
\label{eq:posteriorVarMonotonic}
\end{equation}

\section{IBM Q Error Rates \textit{\&} Calibration Data}\label{app:caldata}
Here, we tabulate the error rate and device calibration data for experiments run on \texttt{ibmq\_montreal} and \texttt{ibmq\_algiers}. The data from \texttt{ibmq\_montreal} was gathered in 2021, whereas the data from \texttt{ibmq\_algiers} was gathered in 2024.

\newpage
\onecolumngrid

\begin{table*}[h!]
\begin{tabular}{|l|l|l|l|l|l|l|l|l|}
\cline{1-9}
qubit & $T_1$ (us) & $T_2$ (us) & frequency (GHz)& anharmonicity  (GHz) & readout error & Pr(prep 1, measure 0)& Pr(prep 0, measure 1)& readout length  (ns) \\\cline{1-9}
Q0 & 108.705 & 71.796 & 4.911 & -0.340 & 0.012 & 0.016 & 0.008 & 5201.778 \\\cline{1-9}
Q1 & 70.072 & 21.384 & 4.835 & -0.324 & 0.014 & 0.019 & 0.010 & 5201.778 \\\cline{1-9}
Q2 & 71.356 & 132.017 & 4.982 & -0.340 & 0.013 & 0.018 & 0.008 & 5201.778 \\\cline{1-9}
Q3 & 106.089 & 83.643 & 5.105 & -0.335 & 0.016 & 0.024 & 0.008 & 5201.778 \\\cline{1-9}
Q4 & 98.017 & 138.303 & 5.004 & -0.338 & 0.013 & 0.016 & 0.009 & 5201.778 \\\cline{1-9}
Q5 & 109.197 & 101.965 & 5.033 & -0.337 & 0.023 & 0.024 & 0.022 & 5201.778 \\\cline{1-9}
Q6 & 214.215 & 24.553 & 4.951 & -0.390 & 0.104 & 0.088 & 0.119 & 5201.778 \\\cline{1-9}
Q7 & 121.456 & 88.514 & 4.906 & -0.323 & 0.234 & 0.230 & 0.238 & 5201.778 \\\cline{1-9}
Q8 & 131.283 & 58.738 & 4.908 & -0.324 & 0.014 & 0.018 & 0.010 & 5201.778 \\\cline{1-9}
Q9 & 105.284 & 103.325 & 5.045 & -0.338 & 0.047 & 0.089 & 0.006 & 5201.778 \\\cline{1-9}
Q10 & 124.091 & 75.809 & 5.082 & -0.337 & 0.012 & 0.020 & 0.004 & 5201.778 \\\cline{1-9}
Q11 & 112.230 & 39.579 & 5.034 & -0.337 & 0.015 & 0.020 & 0.011 & 5201.778 \\\cline{1-9}
Q12 & 138.071 & 88.803 & 4.972 & -0.322 & 0.014 & 0.023 & 0.006 & 5201.778 \\\cline{1-9}
Q13 & 69.351 & 67.000 & 4.868 & -0.340 & 0.007 & 0.009 & 0.004 & 5201.778 \\\cline{1-9}
Q14 & 110.879 & 99.727 & 4.961 & -0.323 & 0.013 & 0.021 & 0.004 & 5201.778 \\\cline{1-9}
Q15 & 105.287 & 27.802 & 5.034 & -0.338 & 0.028 & 0.045 & 0.011 & 5201.778 \\\cline{1-9}
Q16 & 89.486 & 60.709 & 5.086 & -0.337 & 0.010 & 0.014 & 0.005 & 5201.778 \\\cline{1-9}
Q17 & 119.918 & 141.041 & 5.072 & -0.338 & 0.013 & 0.017 & 0.008 & 5201.778 \\\cline{1-9}
Q18 & 83.044 & 33.400 & 4.981 & -0.326 & 0.026 & 0.031 & 0.021 & 5201.778 \\\cline{1-9}
Q19 & 74.795 & 107.818 & 4.983 & -0.321 & 0.014 & 0.021 & 0.006 & 5201.778 \\\cline{1-9}
Q20 & 97.104 & 134.271 & 5.082 & -0.338 & 0.009 & 0.014 & 0.004 & 5201.778 \\\cline{1-9}
Q21 & 111.643 & 43.007 & 5.074 & -0.308 & 0.023 & 0.028 & 0.018 & 5201.778 \\\cline{1-9}
Q22 & 100.898 & 157.914 & 5.057 & -0.338 & 0.021 & 0.031 & 0.011 & 5201.778 \\\cline{1-9}
Q23 & 112.757 & 61.289 & 4.973 & -0.337 & 0.018 & 0.027 & 0.008 & 5201.778 \\\cline{1-9}
Q24 & 82.669 & 56.943 & 5.052 & -0.338 & 0.179 & 0.263 & 0.096 & 5201.778 \\\cline{1-9}
Q25 & 67.330 & 47.877 & 4.934 & -0.323 & 0.034 & 0.064 & 0.004 & 5201.778 \\\cline{1-9}
Q26 & 79.772 & 106.937 & 5.000 & -0.339 & 0.012 & 0.017 & 0.008 & 5201.778 \\\cline{1-9}
\end{tabular}
\caption{\textbf{IBMQ Montreal Calibration Data.} Calibration data from \texttt{ibmq\_montreal} from the time of our FMB demonstrations.}
    \label{tab:ibmq_montreal_calibration}
\end{table*}

\begin{table*}[h!]
\begin{tabular}{|l|l|l||l|l|l|}
   \cline{1-6}
    qubit & Single Qubit Error $(\%)$ & Gate Length (ns) & qubits & Two Qubit Error $(\%)$ & Gate Length (ns) \\\cline{1-6}
    Q0 & 0.018 & 35.556 & (Q0, Q1) & 0.658 & 384.000  \\\cline{1-6}
    Q1 & 0.026 & 35.556 & (Q1, Q2) & 0.848 & 533.334  \\\cline{1-6}
    Q2 & 0.025 & 35.556 & (Q2, Q3) & 0.718 & 369.778  \\\cline{1-6}
    Q3 & 0.039 & 35.556 & (Q3, Q5) & 0.679 & 362.667  \\\cline{1-6}
    Q4 & 0.021 & 35.556 & (Q4, Q1) & 1.196 & 526.222  \\\cline{1-6}
    Q5 & 0.023 & 35.556 & (Q5, Q8) & 0.683 & 355.556  \\\cline{1-6}
    Q6 & 0.044 & 35.556 & (Q6, Q7) & 2.528 & 490.667  \\\cline{1-6}
    Q7 & 0.043 & 35.556 & (Q7, Q4) & 2.112 & 298.667  \\\cline{1-6}
    Q8 & 0.028 & 35.556 & (Q8, Q9) & 0.964 & 369.778  \\\cline{1-6}
    Q9 & 0.044 & 35.556 & (Q10, Q12) & 0.775 & 376.889\\\cline{1-6}
    Q10 & 0.037 & 35.556 & (Q10, Q7) & 2.298 & 554.667 \\\cline{1-6}
    Q11 & 0.033 & 35.556 & (Q11, Q8) & 1.402 & 448.000 \\\cline{1-6}
    Q12 & 0.025 & 35.556 & (Q12, Q13) & 0.667 & 391.111\\\cline{1-6}
    Q13 & 0.018 & 35.556 & (Q13, Q14) & 0.800 & 490.667\\\cline{1-6}
    Q14 & 0.027 & 35.556 & (Q14, Q11) & 0.664 & 341.334 \\\cline{1-6}
    Q15 & 0.076 & 35.556 & (Q15, Q12) & 1.420 & 369.778\\\cline{1-6}
    Q16 & 0.083 & 35.556 & (Q16, Q14) & 1.000 & 320.000 \\\cline{1-6}
    Q17 & 0.031 & 35.556 & (Q17, Q18) & 0.795 & 348.444\\\cline{1-6}
    Q18 & 0.031 & 35.556 & (Q18, Q21) & 1.164 & 405.334 \\\cline{1-6}
    Q19 & 0.029 & 35.556 & (Q19, Q16) & 0.957 & 270.222\\\cline{1-6}
    Q20 & 0.041 & 35.556 & (Q20, Q19) & 0.737 & 320.000\\\cline{1-6}
    Q21 & 0.044 & 35.556 & (Q21, Q23) & 0.858 & 391.111\\\cline{1-6}
    Q22 & 0.025 & 35.556 & (Q22, Q19) & 0.696 & 291.556\\\cline{1-6}
    Q23 & 0.029 & 35.556 & (Q23, Q24) & 2.167 & 405.334 \\\cline{1-6}
    Q24 & 1.613 & 35.556 & (Q24, Q25) & 1.134 & 376.889\\\cline{1-6}
    Q25 & 0.029 & 35.556 & (Q25, Q22) & 1.241 & 576.000\\\cline{1-6}
    Q26 & 0.024 & 35.556 & (Q26, Q25) & 0.815 & 369.778\\\cline{1-6}
    & & & (Q18, Q15) & 1.848 & 597.334  \\\cline{1-6}
\end{tabular}
\caption{\textbf{IBMQ Montreal Qubit Error Rates.} Single and two qubit error rates from\texttt{ibmq\_montreal} from the time of our FMB demonstrations.}
\end{table*}

\begin{table*}[h!]
\begin{tabular}{|l|l|l|l|l|l|l|l|l|}
\cline{1-9}
qubit & $T_1$ (us) & $T_2$ (us) & frequency (GHz)& anharmonicity  (GHz) & readout error & Pr(prep 1, measure 0)& Pr(prep 0, measure 1)& readout length  (ns) \\\cline{1-9}
Q0 & 77.724 & 111.327 & 5.071 & -0.328 & 0.019 & 0.030 & 0.007 & 3512.889 \\\cline{1-9}
Q1 & 128.570 & 234.468 & 4.930 & -0.331 & 0.014 & 0.021 & 0.007 & 3512.889 \\\cline{1-9}
Q2 & 157.032 & 202.960 & 4.670 & -0.337 & 0.015 & 0.022 & 0.007 & 3512.889 \\\cline{1-9}
Q3 & 4.993 & 18.456 & 4.889 & -0.331 & 0.012 & 0.020 & 0.005 & 3512.889 \\\cline{1-9}
Q4 & 97.147 & 61.235 & 5.021 & -0.330 & 0.021 & 0.026 & 0.016 & 3512.889 \\\cline{1-9}
Q5 & 106.113 & 210.447 & 4.970 & -0.330 & 0.018 & 0.028 & 0.007 & 3512.889 \\\cline{1-9}
Q6 & 94.729 & 76.030 & 4.966 & -0.329 & 0.024 & 0.042 & 0.006 & 3512.889 \\\cline{1-9}
Q7 & 87.569 & 138.928 & 4.894 & -0.331 & 0.026 & 0.040 & 0.011 & 3512.889 \\\cline{1-9}
Q8 & 210.779 & 249.220 & 4.792 & -0.333 & 0.022 & 0.025 & 0.019 & 3512.889 \\\cline{1-9}
Q9 & 88.344 & 116.261 & 4.955 & -0.331 & 0.012 & 0.017 & 0.007 & 3512.889 \\\cline{1-9}
Q10 & 106.135 & 275.817 & 4.959 & -0.331 & 0.038 & 0.063 & 0.013 & 3512.889 \\\cline{1-9}
Q11 & 169.838 & 259.322 & 4.666 & -0.333 & 0.037 & 0.037 & 0.037 & 3512.889 \\\cline{1-9}
Q12 & 147.982 & 261.975 & 4.743 & -0.333 & 0.015 & 0.019 & 0.011 & 3512.889 \\\cline{1-9}
Q13 & 101.292 & 239.575 & 4.889 & -0.328 & 0.013 & 0.017 & 0.008 & 3512.889 \\\cline{1-9}
Q14 & 144.048 & 184.800 & 4.780 & -0.333 & 0.040 & 0.051 & 0.029 & 3512.889 \\\cline{1-9}
Q15 & 152.841 & 105.949 & 4.858 & -0.333 & 0.021 & 0.028 & 0.014 & 3512.889 \\\cline{1-9}
Q16 & 87.767 & 194.604 & 4.980 & -0.330 & 0.010 & 0.017 & 0.004 & 3512.889 \\\cline{1-9}
Q17 & 43.440 & 151.820 & 5.003 & -0.330 & 0.015 & 0.021 & 0.009 & 3512.889 \\\cline{1-9}
Q18 & 185.467 & 307.095 & 4.781 & -0.333 & 0.076 & 0.089 & 0.062 & 3512.889 \\\cline{1-9}
Q19 & 121.035 & 216.184 & 4.810 & -0.332 & 0.036 & 0.040 & 0.031 & 3512.889 \\\cline{1-9}
Q20 & 95.722 & 176.897 & 5.048 & -0.328 & 0.023 & 0.033 & 0.013 & 3512.889 \\\cline{1-9}
Q21 & 139.250 & 214.520 & 4.943 & -0.331 & 0.019 & 0.027 & 0.010 & 3512.889 \\\cline{1-9}
Q22 & 119.703 & 157.558 & 4.911 & -0.332 & 0.026 & 0.044 & 0.007 & 3512.889 \\\cline{1-9}
Q23 & 90.834 & 173.980 & 4.893 & -0.332 & 0.048 & 0.060 & 0.035 & 3512.889 \\\cline{1-9}
Q24 & 135.594 & 41.362 & 4.671 & -0.336 & 0.024 & 0.034 & 0.014 & 3512.889 \\\cline{1-9}
Q25 & 80.640 & 193.740 & 4.759 & -0.334 & 0.014 & 0.019 & 0.009 & 3512.889 \\\cline{1-9}
Q26 & 103.393 & 173.879 & 4.954 & -0.330 & 0.014 & 0.019 & 0.008 & 3512.889 \\\cline{1-9}
\end{tabular}
\caption{\textbf{IBMQ Algiers Calibration Data.} Calibration data from \texttt{ibmq\_algiers} from the time of our FMB demonstrations.}
    \label{tab:ibmq_algiers_calibration}
\end{table*}

\begin{table*}[h!]
\begin{tabular}{|l|l|l||l|l|l|}
    \cline{1-6}
    qubit & Single Qubit Error $(\%)$ & Gate Length (ns) & qubits & Two Qubit Error $(\%)$ & Gate Length (ns) \\\cline{1-6}
    Q0 & 0.033 & 35.560 & (Q0, Q1) & 0.727 & 419.560 \\\cline{1-6}
    Q1 & 0.020 & 35.560 & (Q1, Q2) & 0.896 & 704.000 \\\cline{1-6}
    Q2 & 0.016 & 35.560 & (Q2, Q3) & 0.557 & 376.890 \\\cline{1-6}
    Q3 & 0.038 & 35.560 & (Q3, Q5) & 1.037 & 426.670 \\\cline{1-6}
    Q4 & 0.029 & 35.560 & (Q4, Q1) & 0.519 & 312.890 \\\cline{1-6}
    Q5 & 0.021 & 35.560 & (Q5, Q8) & 100.000 & 256.000 \\\cline{1-6}
    Q6 & 0.024 & 35.560 & (Q6, Q7) & 1.193 & 248.890 \\\cline{1-6}
    Q7 & 0.032 & 35.560 & (Q7, Q4) & 0.982 & 604.440 \\\cline{1-6}
    Q8 & 0.017 & 35.560 & (Q8, Q9) & 0.915 & 604.440 \\\cline{1-6}
    Q9 & 0.057 & 35.560 & (Q10, Q12) & 0.707 & 604.440 \\\cline{1-6}
    Q10 & 0.025 & 35.560 & (Q10, Q7) & 0.780 & 398.220 \\\cline{1-6}
    Q11 & 0.018 & 35.560 & (Q11, Q8) & 0.652 & 604.440 \\\cline{1-6}
    Q12 & 0.019 & 35.560 & (Q12, Q13) & 0.563 & 604.440 \\\cline{1-6}
    Q13 & 0.015 & 35.560 & (Q13, Q14) & 0.690 & 604.440 \\\cline{1-6}
    Q14 & 0.019 & 35.560 & (Q14, Q11) & 0.414 & 391.110 \\\cline{1-6}
    Q15 & 0.018 & 35.560 & (Q15, Q12) & 0.496 & 369.780 \\\cline{1-6}
    Q16 & 0.019 & 35.560 & (Q16, Q14) & 0.630 & 291.560 \\\cline{1-6}
    Q17 & 0.024 & 35.560 & (Q17, Q18) & 100.000 & 248.890 \\\cline{1-6}
    Q18 & 0.021 & 35.560 & (Q18, Q21) & 0.642 & 497.780 \\\cline{1-6}
    Q19 & 0.019 & 35.560 & (Q19, Q16) & 1.082 & 682.670 \\\cline{1-6}
    Q20 & 0.038 & 35.560 & (Q20, Q19) & 0.670 & 369.780 \\\cline{1-6}
    Q21 & 0.030 & 35.560 & (Q21, Q23) & 0.802 & 362.670 \\\cline{1-6}
    Q22 & 0.075 & 35.560 & (Q22, Q19) & 0.738 & 327.110 \\\cline{1-6}
    Q23 & 0.029 & 35.560 & (Q23, Q24) & 0.988 & 604.440 \\\cline{1-6}
    Q24 & 0.016 & 35.560 & (Q24, Q25) & 0.549 & 433.780 \\\cline{1-6}
    Q25 & 0.016 & 35.560 & (Q25, Q22) & 1.219 & 448.000 \\\cline{1-6}
    Q26 & 0.014 & 35.560 & (Q26, Q25) & 1.016 & 312.890 \\\cline{1-6}
    & & & (Q15, Q18) & 0.525 & 305.780 \\\cline{1-6}
\end{tabular}
\caption{\textbf{IBMQ Algiers Qubit Error Rates.} Single and two qubit error rates from \texttt{ibmq\_algiers} from the time of our FMB demonstrations.}
\end{table*}

\end{document}